\documentclass{elsarticle}
\pdfoutput=1 

\usepackage[a4paper, top=1.0in, bottom=1.2in, left=1.0in, right=1.0in]{geometry}
\usepackage[utf8]{inputenc}
\usepackage[tbtags]{amsmath} 
\usepackage{latexsym}

\usepackage{stix2}

\usepackage{framed}
\usepackage{multirow}
\usepackage{placeins}
\usepackage{url}
\usepackage{xcolor}
\definecolor{newcolor}{rgb}{.8,.349,.1}
\usepackage{booktabs} 
\usepackage{empheq}

\usepackage{hyperref} 

\newdefinition{rmk}{Remark}

\newcommand{\ie}{i.e.\ }
\newcommand{\eg}{e.g.\ }

\usepackage{tikz} 
\usetikzlibrary{shapes,arrows,arrows.meta,matrix,decorations.pathmorphing,shapes.geometric,positioning,calc}  

\def\centerarc[#1](#2)(#3:#4:#5){ \draw[#1] ($(#2)+({#5*cos(#3)},{#5*sin(#3)})$) arc (#3:#4:#5); }

\renewcommand{\v}[1]{\ensuremath{\mathbf{#1}}} 

\renewcommand{\d}[2]{\frac{\mathrm{d} #1}{\mathrm{d} #2}} 
\newcommand{\dd}[2]{\frac{\mathrm{d}^2 #1}{\mathrm{d} #2^2}} 
\newcommand{\ddd}[2]{\frac{\mathrm{d}^3 #1}{\mathrm{d} #2^3}} 
\newcommand{\pd}[2]{\frac{\partial #1}{\partial #2}} 
\newcommand{\pdd}[2]{\frac{\partial^2 #1}{\partial #2^2}} 

\newcommand{\dotp}[2]{\langle #1 \, {,} \, #2 \rangle} 

\usepackage{stmaryrd} 

\usepackage[english]{babel} 
\usepackage[titletoc,title]{appendix}

\begin{document}

\title{Energy-conserving formulation of the two-fluid model for incompressible two-phase flow in channels and pipes}              
\author[add1,add2]{J.F.H. Buist}
\ead{jurriaan.buist@cwi.nl, corresponding author}
\author[add1]{B. Sanderse}
\author[add3]{S. Dubinkina}
\author[add2,add4]{R.A.W.M. Henkes}
\author[add5]{C.W. Oosterlee}

\address[add1]{Centrum Wiskunde \& Informatica (CWI), Amsterdam, The Netherlands}
\address[add2]{Delft University of Technology, Delft, The Netherlands}
\address[add3]{Vrije Universiteit Amsterdam, Amsterdam, The Netherlands}
\address[add4]{Shell Technology Centre Amsterdam, Amsterdam, The Netherlands}
\address[add5]{Utrecht University, Utrecht, The Netherlands}

\begin{abstract}

We show that the one-dimensional (1D) two-fluid model (TFM) for stratified flow in channels and pipes (in its incompressible, isothermal form) satisfies an energy conservation equation, which arises naturally from the mass and momentum conservation equations that constitute the model. 
This result extends upon earlier work on the shallow water equations (SWE), with the important difference that we include non-conservative pressure terms in the analysis, and that we propose a formulation that holds for ducts with an arbitrary cross-sectional shape, with the 2D channel and circular pipe geometries as special cases. 

The second novel result of this work is the formulation of a finite volume scheme for the TFM that satisfies a discrete form of the continuous energy equation. 
This discretization is derived in a manner that runs parallel to the continuous analysis. 
Due to the non-conservative pressure terms it is essential to employ a staggered grid, which requires careful consideration in defining the discrete energy and energy fluxes, and the relations between them and the discrete model.
Numerical simulations confirm that the discrete energy is conserved.

\end{abstract}



\maketitle



\section{Introduction}

The one-dimensional (1D) two-fluid model (TFM) is a dynamic model for stratified flow in channels and pipes.
It simplifies the full three-dimensional multiphase flow problem by resolving only the cross-sectionally averaged quantities (hold-ups, velocities, and pressure), which are often of practical interest.
There are many variants of the model, but the basic idea, of two interacting fluids whose behaviour is cross-sectionally averaged to obtain a 1D model, was introduced by Wallis (1969) \cite{Wallis1969} and Ishii (1975) \cite{Ishii1975}. 
The model has among others applications in the oil and gas industry \cite{GoldszalDanielsonBansalEtAl2007}, in CO$_{2}$ transport and storage \cite{AursandHammerMunkejordEtAl2013}, and in nuclear reactor safety analysis \cite{BerryZouZhaoEtAl2014}. 

An unsolved issue with the basic version of the TFM is that the initial value problem for the governing equations is only conditionally well-posed \cite{LyczkowskiGidaspowSolbrigEtAl1978}.
This means that it is well-posed for some flow configurations and ill-posed for others (\eg when there is a large velocity difference between the two fluids).
Conventionally, ill-posedness of the TFM is demonstrated by a linear stability analysis which shows an unbounded growth rate for the smallest wavelengths, when the values of the model variables are such that the eigenvalues are complex.
In this case the solution is said to carry no physical meaning \cite{LiaoMeiKlausner2008}. 
However, when drawing conclusions on the well-posedness of the TFM, it is important to also consider its nonlinear aspects, and not only rely on a linearized analysis 
\cite{KreissYstrom2006, StewartWendroff1984}.
Examples of studies that have included nonlinear effects in the TFM analysis can be found in \cite{Keyfitz2001,LopezdeBertodanoFullmerClausseEtAl2017}.
However, a complete nonlinear analysis, with implications for obtaining a robust discretization, is still missing.

In this work, we strive towards such a nonlinear analysis by presenting an expression for an energy which is conserved by the full (nonlinear) TFM, in its incompressible and isothermal form.
This approach is motivated by the fact that for the incompressible Navier-Stokes equations such an analysis provides stability estimates \cite{CoppolaCapuanodeLuca2019,Sanderse2013b}, and that for compressible equations it is closely related to the concept of entropy stability \cite{Tadmor2003}. Important to note is that such an energy is not the thermodynamic energy for which a separate conservation equation exists in the compressible TFM.
Rather, the considered energy conservation is an inherent property of the mass and momentum conservation equations that constitute the incompressible TFM: the energy is a secondary conserved quantity of the model.
Its physical meaning is therefore the mechanical energy of the system (kinetic plus potential energy).

In order to derive this mechanical energy equation, we take the approach from \cite{FjordholmMishraTadmor2011}, in which the dot product of the shallow water equations (SWE) and a vector of entropy variables is taken in such a way that a scalar energy equation results. However, an important difference with the SWE (and two-layer SWE \cite{Fjordholm2012}) is the presence of non-conservative pressure terms that are linked to the constraint that the fluid phases have to fill the cross section. 
Another important difference is that we consider arbitrary duct geometries, as opposed to the 1D SWE which in effect utilizes a planar channel geometry. Given these differences, the key challenge is thus to find a conserved energy and corresponding energy flux function for the TFM, and this will be the first main focus of this paper.

The second focus of this paper is to derive a spatial discretization which conserves a discrete version of the energy.
Again, our approach is inspired by methods which have been developed for the SWE \cite{FjordholmMishraTadmor2009}.
An important difference is that these methods are designed for collocated grids, while we will adapt them to a staggered grid.
This is motivated by the presence of the (non-conservative) pressure terms in the TFM, which makes the use of a staggered grid much more convenient (similar to the case of the incompressible Navier-Stokes equations \cite{CoppolaCapuanodeLuca2019}). 
However, the staggered grid introduces new challenges, for example in terms of  the definitions of the energy and energy fluxes. 
We will derive a discretization method that tackles these issues and propose a new set of  numerical fluxes on a staggered grid that are energy conservative.
This discretization can also be viewed as an extension of the SWE discretization found in \cite{vantHofVeldman2012}, where a different method is used to obtain a mass-, \mbox{momentum-,} and energy-conserving discretization on a staggered grid.

This paper is set up as follows.
First, in \autoref{sec:two-fluid_model} we present the governing equations of the TFM.
In \autoref{sec:continuous_energy_conservation} we discuss the conditions for energy conservation, and introduce an energy and energy flux that satisfies these conditions, providing local and global energy conservation equations for the continuous TFM.
We outline how the equations are discretized in \autoref{sec:discretization}, while leaving open the specific form of the numerical fluxes. 
Then, in \autoref{sec:semi-discrete_local_energy_conservation}, we present the discrete versions of the continuous conditions for energy conservation, and propose a set of new conservative numerical fluxes.  
Finally, in \autoref{sec:simulations} we present numerical results which demonstrate exact conservation of the aforementioned energy.

\section{Governing equations}\label{sec:two-fluid_model}
The 1D TFM, as considered in this work, describes the separated flow of a (heavier) lower fluid $L$ and a (lighter) upper fluid $U$ through a channel or pipe. 
It can be derived by applying a cross-sectional averaging procedure to the Navier-Stokes equations \cite{IshiiMishima1984, StewartWendroff1984}.
An important assumption made in the derivation of the model is that the streamwise length scale is much larger than the normal length scale (\ie the pipe diameter), which is referred to as the long wavelength assumption. 
As a consequence, along the normal direction the flow is in hydrostatic balance.
We will omit source terms, such as wall friction, since such terms are sources or sinks of energy, and we are interested in the energy conservation properties of the core model.
Good discussions of the assumptions underlying the TFM are given by \cite{Montini2011, Munkejord2006}. 

\begin{figure}[thb]
\large \centering
\begin{tikzpicture}[scale=1.5,baseline=(current bounding box.north)]


\filldraw[fill=blue!30!white, draw=black] (-3,0) sin (-1.5,0.25) cos (0,0) sin (1.5,-0.25) cos (3,0) -- (3,-1) -- (-3,-1) -- cycle;

\draw[-{Latex[width = 2.2mm, length = 2.2mm]},dashed] (-3.2, -1.075)--(-3.2,0.5);
\node [left] at (-3.2,0.3) {$h$};
\draw[-{Latex[width = 2.2mm, length = 2.2mm]}] (3.25, 0)--(3.25,-1);
\draw[-{Latex[width = 2.2mm, length = 2.2mm]}] (3.25, 0)--(3.25,1);
\node [right] at (3.25,0) {$H$};

\draw[-{Latex[width = 2.2mm, length = 2.2mm]},dashed] (-3.07, -1.2)--(1,-1.2);
\node [right] at (1,-1.2) {$s$};

\draw[] (-3.2,-1.2) circle (0.13);
\draw[] (-3.292,-1.108) -- (-3.108,-1.292);
\draw[] (-3.108,-1.108) -- (-3.292,-1.292);

\draw[thick] (-3,-1) rectangle (3,1);

\draw[-{Latex[width = 2.2mm, length = 2.2mm]}] (-0.5, -0.5)--(0.5,-0.5);
\node [above] at (0,-0.5) {$u_L$};

\draw[-{Latex[width = 2.2mm, length = 2.2mm]}] (-0.5, 0.5)--(0.5,0.5);
\node [above] at (0,0.5) {$u_U$};




\draw[-{Latex[width = 2.2mm, length = 2.2mm]}] (1.5, -0.625)--(1.5,-0.25);
\draw[-{Latex[width = 2.2mm, length = 2.2mm]}] (1.5, -0.625)--(1.5,-1);
\node [right] at (1.5,-0.625) {$H_L$};

\draw[-{Latex[width = 2.2mm, length = 2.2mm]}] (1.5, 0.375)--(1.5,-0.25);
\draw[-{Latex[width = 2.2mm, length = 2.2mm]}] (1.5, 0.375)--(1.5,1);
\node [right] at (1.5, 0.4) {$H_U$};






\filldraw[fill=blue!30!white, draw=black] (-5.9798,-0.2) -- (-4.0202,-0.2) arc (-11.087:-168.913:1) -- cycle;

\draw[thick] (-5,0) circle (1);

\node at (-5,-0.6) {$A_L$};
\node at (-5,0.6) {$A_U$};

\centerarc[-{Latex[width = 2.2mm, length = 2.2mm]}](-5,0)(-90:-168.913:1.12);
\centerarc[-{Latex[width = 2.2mm, length = 2.2mm]}](-5,0)(-90:-11.087:1.12);
\node[right] at (-4.15,-0.73) {$P_L$};

\centerarc[-{Latex[width = 2.2mm, length = 2.2mm]}](-5,0)(90:191.087:1.12);
\centerarc[-{Latex[width = 2.2mm, length = 2.2mm]}](-5,0)(90:-11.087:1.12);
\node[right] at (-4.15,0.73) {$P_U$};

\draw[-{Latex[width = 2.2mm, length = 2.2mm]}](-5,-0.1) -- (-4.0202,-0.1);
\draw[-{Latex[width = 2.2mm, length = 2.2mm]}](-5,-0.1) -- (-5.9798,-0.1) ;
\node[above] at (-5,-0.1) {$P_\mathrm{int}$};
\end{tikzpicture}%
 \normalsize 
\caption{A schematic of stratified two-fluid flow in ducts (a circular pipe segment is shown as an example) described by the 1D TFM.}
\label{fig:two-fluid_schematic}
\end{figure}
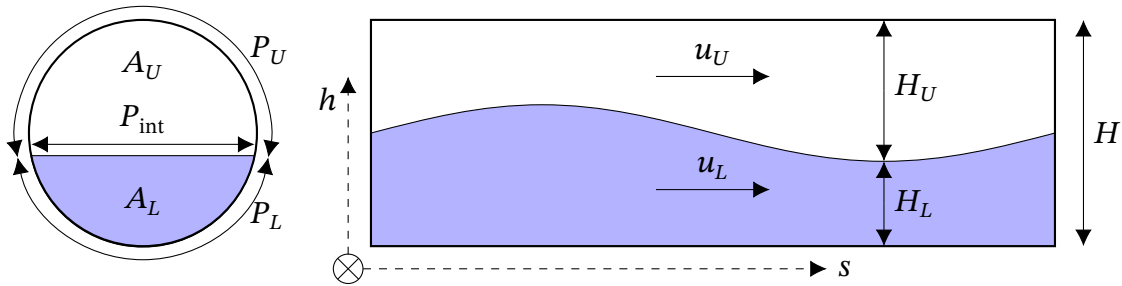

The cross-sectionally averaged equations can be written in the following concise form \cite{SanderseSmithHendrix2017,SanderseVeldman2019} (with $\v{q} = \v{q}(s,t)$):
\begin{equation}
\pd{\v{q}}{t}  + \pd{\v{f}( \v{q} )}{s}  +  \v{d}(\v{q}) \pd{p}{s} = \v{0},
\label{eq:governing_equations/conservative_v1} 
\end{equation}
where $\v{q}$ constitutes the vector of `conserved' variables\footnote{Note that the pressure term is not in conservative form, so $q_{3}$ and $q_{4}$ are individually not conserved, but $q_3+q_4$ is.}, namely the mass and momentum of each phase:
\begin{equation} 
\v{q}^T = 
\begin{bmatrix}
q_1 &
q_2 &
q_3 &
q_4
\end{bmatrix}
=
\begin{bmatrix}
\rho_U A_U &
\rho_L A_L &
\rho_U u_U A_U &
\rho_L u_L A_L
\end{bmatrix}.
\label{eq:governing_equations/conservative_v1/solution_vector}
\end{equation}
Here $\rho_U$ and $\rho_L$ are the densities, $A_U$ and $A_L$ are the cross sections, and $u_U$ and $u_L$ are the averaged velocities, all of the upper and lower fluids,  respectively. 
We consider the isothermal, incompressible case, so that $\rho_U$ and $\rho_L$ are constant. 

The fluxes $\v{f}$ describe convection of mass and momentum and gradients in the interface level. In terms of $\v{q}$ they are given by
\begin{equation}
\begin{split}
\v{f}(\v{q})^T &=
\begin{bmatrix}
q_3 &
q_4 &
\frac{q_3^2}{q_1} -  \rho_U g_n  \widehat{H}_U &
\frac{q_4^2}{q_2} - \rho_L g_n  \widehat{H}_L
\end{bmatrix} \\
&=
\begin{bmatrix}
\rho_U u_U A_U &
\rho_L u_L A_L &
\rho_U u_U^2 A_U -  \rho_U g_n  \widehat{H}_U &
\rho_L u_L^2 A_L - \rho_L g_n  \widehat{H}_L
\end{bmatrix},
\end{split}
\label{eq:governing_equations/conservative_v1/flux_vector}
\end{equation}
where $\widehat{H}_U = \widehat{H}_U(\v{q})$ and $\widehat{H}_L = \widehat{H}_L(\v{q})$ are geometric terms (to be discussed shortly), and $g_n$ is the gravitational acceleration in the normal direction.

The fifth variable is the interface pressure $p$, and the non-conservative pressure terms are given by $\v{d} (\partial p/\partial s)$ with
\begin{equation}
\v{d}(\v{q})^T = 
\begin{bmatrix}
0 &
0 &
 \frac{q_1}{\rho_U}  &
 \frac{q_2}{\rho_L} 
\end{bmatrix}
=
\begin{bmatrix}
0 &
0 &
A_U  &
A_L 
\end{bmatrix}.
\label{eq:governing_equations/conservative_v1/pressure_vector}
\end{equation}

The quantities $\widehat{H}_U = \widehat{H}_U(A_U(q_1,\rho_U))$ and $\widehat{H}_L = \widehat{H}_L(A_L(q_2,\rho_L))$ are geometry-dependent and are defined by
\begin{equation}
\widehat{H}_U \coloneqq  \int_{a_U}   (h-H_L) \, \mathrm{d} a, \quad \quad
\widehat{H}_L \coloneqq   \int_{a_L}   (h-H_L) \, \mathrm{d} a.
\label{eq:governing_equations/level_gradient_integrals}
\end{equation}
Here the difference between the coordinate $h$ and the two-fluid interface height $H_L$ is integrated over the area $a_U$ occupied by the upper fluid and the area $a_L$ occupied by the lower fluid, respectively.
Using these general expressions, the model equations are valid for arbitrarily shaped cross sections. 
See \autoref{sec:geometric_details} for evaluations of the integrals for the 2D channel and circular pipe geometries.
The spatial derivatives of $\widehat{H}_U$ and $\widehat{H}_L$ that appear in the fluxes $\v{f}$ are known as the level gradient terms, which result from the hydrostatic variation of the pressure.

Since the upper and lower fluid together fill the pipe, the system is subject to the volume constraint 
\begin{equation}
\frac{q_1}{\rho_U} + \frac{q_2}{\rho_L} = A.
\label{eq:governing_equations/volume_constraint}
\end{equation}
The entire system therefore consists of four evolution equations plus one constraint, and four `conserved' variables plus the pressure. In our incompressible setting, a derived constraint can be obtained by 
differentiating the constraint \eqref{eq:governing_equations/volume_constraint} and substituting the mass equations, leading to \cite{SanderseVeldman2019}:
 \begin{equation}
 \pd{}{s} \left( \frac{q_3}{\rho_U} + \frac{q_4}{\rho_L} \right) = 0,
 \label{eq:governing_equations/volumetric_flux_constraint_diff}
 \end{equation}
which can be integrated in space to give that the volumetric flow $Q$ is constant in space, and a function of time only:
  \begin{equation}
 Q(\v{q}) \coloneqq  \frac{q_3}{\rho_U} + \frac{q_4}{\rho_L} = Q(t).
 \label{eq:governing_equations/volumetric_flux_constraint}
 \end{equation}
This derived constraint, termed the volumetric flow constraint, can be seen as the incompressibility constraint for the TFM.

We can use these constraints to set up an equation for the pressure.   
The pressure equation is obtained by summing the momentum equations \cite{SanderseVeldman2019}:
\begin{equation}
 \v{l}^T \v{d} \pd{p}{s} = - \v{l}^T \left( \pd{\v{q} }{t} + \pd{\v{f}}{s}  \right), \quad \text{with} \quad 
 \v{l}^T = 
\begin{bmatrix}
0 & 0 & \frac{1}{\rho_U} & \frac{1}{\rho_L}
\end{bmatrix},
\end{equation}
which can be expanded and rewritten with the definition of $Q$ to yield
 \begin{equation}
 \left(\frac{q_1}{\rho_U^2} + \frac{q_2}{\rho_L^2} \right) \pd{p}{s} = -\d{Q}{t}  - \pd{}{s} \left( \frac{f_3}{\rho_U} + \frac{f_4}{\rho_L} \right). 
 \label{eq:governing_equations/pressure_Poisson_pre_step}
 \end{equation}
Finally, taking the derivative of this equation to $s$ and applying constraint  \eqref{eq:governing_equations/volumetric_flux_constraint_diff}  gives
\begin{equation} 
 \pd{}{s} \left( \left(\frac{q_1}{\rho_U^2} + \frac{q_2}{\rho_L^2} \right) \pd{p}{s} \right) = -\pdd{}{s} \left( \frac{f_3}{\rho_U} + \frac{f_4}{\rho_L} \right).
 \label{eq:governing_equations/pressure_Poisson}
 \end{equation}
This is a `Poisson-type' equation for the pressure, which can be used in place of \eqref{eq:governing_equations/volume_constraint} to close the system of equations. 
In our numerical algorithm (discussed in \autoref{sec:discretization}) we apply a discrete version of \eqref{eq:governing_equations/pressure_Poisson} in this manner.

\section{Energy conservation equation for the continuous two-fluid model}
\label{sec:continuous_energy_conservation}

\subsection{Outline: conditions for energy conservation}

Having set-up the TFM governing equations, the first key objective of this paper is to prove \textit{local} and \textit{global} energy equalities that are \textit{implied} by this equation set. This is similar to the energy analyses for \eg the incompressible Navier-Stokes equations \cite{CoppolaCapuanodeLuca2019}, the SWE \cite{TadmorZhong2008}, and the two-layer SWE \cite{Fjordholm2012}. In all these models, no energy conservation equation is included in the model, but energy conservation follows from the mass and momentum conservation equations alone. 
It can therefore be said that the energy is a secondary conserved quantity.

Our proof of global energy conservation follows the approach in  \cite{FjordholmMishraTadmor2009,FjordholmMishraTadmor2011} and starts by showing that a local energy conservation equation of the form
\begin{equation}
\pd{e}{t} + \pd{}{s} \left( h+j \right) = 0
\label{eq:continuous_energy_conservation/continuous_energy_equation_v3}
\end{equation}
can be derived, purely based on manipulating the governing equations \eqref{eq:governing_equations/conservative_v1}. Here $e(\v{q})$ is the local energy, and $h(\v{q})$ and $j(\v{q})$ are energy fluxes (to be detailed later); $h(\v{q})$ is not be confused with the normal coordinate $h$ shown in \autoref{fig:two-fluid_schematic}. If \eqref{eq:continuous_energy_conservation/continuous_energy_equation_v3} holds, then it can be integrated in space to yield
\begin{align}
\d{E}{t} &= -\left[ h+ j \right]_{s_1}^{s_2} = 0,
\label{eq:continuous_energy_conservation/continuous_energy_equation_v4}
\end{align}
where the last equality (`=0') holds in case of periodic or closed boundaries, and the global energy $E(t)$ is defined as
\begin{equation}
E(t) = \int_{s_1}^{s_2} e \, \mathrm{d}{s}.
\label{eq:continuous_energy_conservation/definition}
\end{equation}


The key is therefore to obtain the local energy conservation equation \eqref{eq:continuous_energy_conservation/continuous_energy_equation_v3}. To achieve this, one first postulates an energy $e(\v{q})$ (typically guided by physical considerations). Second, one calculates the vector of so-called entropy variables, defined as\footnote{We take the convention that $\partial e/\partial \v{q}$ is a row vector, making $\v{v}$ a column vector.}
\begin{equation*}
\v{v}(\v{q}) \coloneqq \left[\pd{e}{\v{q}}\right]^T.
\end{equation*}
Taking the dot product of the system \eqref{eq:governing_equations/conservative_v1} with $\v{v}$ leads to 
\begin{equation}
\dotp{ \v{v} }{ \pd{\v{q}}{t} } +\dotp{ \v{v} }{  \pd{ \v{f}}{s} }  + \dotp{ \v{v} }{  \v{d} \pd{p}{s}} = 0,
\label{eq:continuous_energy_conservation/continuous_energy_equation_v1}
\end{equation}
in which we have ignored source terms (as indicated before), and the brackets denote a dot product over the vector elements:
\begin{equation*}
\dotp{ \v{x} }{ \v{y} } \coloneqq \v{x}^T \v{y}. 
\end{equation*}
The time derivative term can be written as
\begin{equation}
\dotp{ \v{v} }{ \pd{\v{q}}{t}  } = \left( \pd{e}{\v{q}} \right) \pd{\v{q}}{t} = \pd{e}{t},
\label{eq:continuous_energy_conservation/energy_substitution}
\end{equation}
so \eqref{eq:continuous_energy_conservation/continuous_energy_equation_v1} becomes an equation for the time evolution of the energy. 

Given an expression for $e$, the art is to find an energy flux $h$ that satisfies
\begin{equation}
\dotp{ \v{v} }{ \pd{\v{f}}{s} } = \pd{h}{s},
\label{eq:continuous_energy_conservation/h_to_f_condition_basic}
\end{equation}
since then the second term in \eqref{eq:continuous_energy_conservation/continuous_energy_equation_v1} can be written in the (locally) conservative form given by \eqref{eq:continuous_energy_conservation/continuous_energy_equation_v3}. In order to get a condition solely referring to the relations between different functions of $\v{q}$ (\ie independent of $s$), the chain rule (valid for smooth solutions) is employed to convert \eqref{eq:continuous_energy_conservation/h_to_f_condition_basic} to:
\begin{align}
\dotp{ \v{v} }{\pd{\v{f}}{\v{q}} } &= \pd{h}{\v{q}}. \label{eq:continuous_energy_conservation/h_to_f_condition}
\end{align}
This is the condition encountered in \eg \cite{FjordholmMishraTadmor2009} and \cite{FjordholmMishraTadmor2011} for an energy flux $h$ to conserve a given energy $e$ (or, more generally: entropy function) of the SWE. 

Likewise, we need to find a flux $j$ such that the product of $\v{v}$ and the pressure gradient can be written in conservative form:
\begin{equation}
\dotp{\v{v}}{\v{d} \pd{p}{s}} = \pd{j}{s}.
 \label{eq:continuous_energy_conservation/j_condition_basic}
\end{equation}
The difference between $h$ and $j$ lies in the fact that $h$ is responsible for the spatially conservative terms of the governing equations, whereas $j$ takes the non-conservative part into account. Perhaps surprisingly, we will show that these non-conservative terms $\v{d} (\partial p/\partial s)$ can indeed be written in conservative form in the energy equation. 
An alternative formulation of condition \eqref{eq:continuous_energy_conservation/j_condition_basic} is given by
\begin{equation}
\pd{}{s} \left( \dotp{ \v{v} }{\v{d}} p \right) - p \pd{}{s} \dotp{\v{v} }{ \v{d} }= \pd{j}{s}.
 \label{eq:continuous_energy_conservation/j_condition}
\end{equation}
In order for the local energy to be conserved, there must exist a $j$ (for the given $e$ and resulting $\v{v}$) such that this condition is satisfied. 

An important difference between this derivation and the derivation for the SWE as found in \eg \cite{FjordholmMishraTadmor2011} is the non-conservative pressure term. Although the two-layer SWE \cite{AbgrallKarni2009} also features a  non-conservative term, in the TFM the non-conservative term depends on a variable for which there is no evolution equation (namely the pressure).
This pressure term is instead linked directly to the volume constraint \eqref{eq:governing_equations/volume_constraint} and volumetric flow constraint \eqref{eq:governing_equations/volumetric_flux_constraint} \cite{SanderseVeldman2019}, which are not present in the SWE.
For a system in conservative form without source terms, \eqref{eq:continuous_energy_conservation/h_to_f_condition} is the only condition.
This condition is emphasized in literature (\eg \cite{Leveque2002}) as the condition for the existence of an entropy function. 
The derivation of energy conservation for the conservative part of the TFM system thus matches the derivation of an entropy condition for a conservative hyperbolic system.

In summary, the task is to find a set $e$, $h$ and, $j$ which satisfy conditions  \eqref{eq:continuous_energy_conservation/h_to_f_condition_basic} and  \eqref{eq:continuous_energy_conservation/j_condition_basic} for the current model with flux $\v{f}$ and pressure terms $\v{d} (\partial p/\partial s)$.
The alternative conditions \eqref{eq:continuous_energy_conservation/h_to_f_condition} and \eqref{eq:continuous_energy_conservation/j_condition} yield results more directly and will therefore be used in the following section. The result is the local energy conservation equation \eqref{eq:continuous_energy_conservation/continuous_energy_equation_v3}, and global energy conservation then follows directly.

\subsection{Choice of energy and energy fluxes}
\label{ssec:continuous_energy_conservation/specific_energy}

We will show that the energy
\begin{align}
\Aboxed{e &=\rho_U g_n \widetilde{H}_U +  \rho_L g_n \widetilde{H}_L + \frac{1}{2}  \frac{q_3^2}{q_1} + \frac{1}{2}  \frac{q_4^2}{q_2} }\label{eq:continuous_energy_conservation/conservative/e_definition_1} \\
&= \rho_U g_n \widetilde{H}_U +  \rho_L g_n \widetilde{H}_L + \frac{1}{2} \rho_U  A_U u_U^2 + \frac{1}{2} \rho_L  A_L u_L^2, \nonumber
\end{align}
is conserved by the TFM (in absence of source terms). Here $ \widetilde{H}_U = \widetilde{H}_U(A_U(q_1,\rho_U))$ represents the center of mass of the upper fluid multiplied by $A_U$ and $\widetilde{H}_L = \widetilde{H}_L (A_L(q_2,\rho_L))$ represents the center of mass of the lower fluid multiplied by $A_L$ (see \autoref{sec:geometric_details}), so that the first two terms can be recognized as the potential energy of the upper and lower fluid, respectively. The third and fourth terms represent the kinetic energy of the upper and lower fluid, respectively. Therefore, this energy $e$ has a clear physical interpretation.

The entropy variables are given by
\begin{equation}
\v{v} = \left[\pd{e}{\v{q}}\right]^T 
=
\begin{bmatrix}
- \frac{1}{2}  \frac{q_3^2}{q_1^2} +  g_n \d{ \widetilde{H}_U}{A_U} \\
- \frac{1}{2} \frac{q_4^2}{q_2^2} +  g_n  \d{ \widetilde{H}_L}{A_L} \\
 \frac{q_3}{q_1} \\
\frac{q_4}{q_2}
\end{bmatrix}
=
\begin{bmatrix}
- \frac{1}{2}  \frac{q_3^2}{q_1^2} +  g_n (H-H_U) \\
- \frac{1}{2} \frac{q_4^2}{q_2^2} +  g_n H_L \\
 \frac{q_3}{q_1} \\
\frac{q_4}{q_2}
\end{bmatrix},
\label{eq:continuous_energy_conservation/v_definition} 
\end{equation}
with $ H_U = H_U(A_U(q_1,\rho_U))$ and $H_L = H_L(A_L(q_2,\rho_L))$ representing the fluid layer thickness of the upper and lower fluids, respectively (see \autoref{sec:geometric_details}).
It is important that in the energy and energy flux terms concerning the upper fluid we use $\widehat{H}_U(A_U)$,  $\widetilde{H}_U(A_U)$, and $H_U(A_U)$, while for the lower fluid we use $\widehat{H}_L(A_L)$,  $\widetilde{H}_L(A_L)$, and $H_L(A_L)$. 
It is possible to use the volume constraint to change this functional dependence, but our choice leads to an elegant form of the energy conservation conditions.

The task is to find $h$ and $j$. We start with $j$: the pressure term needs to satisfy \eqref{eq:continuous_energy_conservation/j_condition}. Straightforward evaluation gives
\begin{equation*}
\dotp{\v{v}}{\v{d}} = \frac{q_3}{q_1} \frac{q_1}{\rho_{U}} + \frac{q_4}{q_2} \frac{q_2}{\rho_{L}} = Q,
\end{equation*}
with $Q$ the volumetric flow rate given by  \eqref{eq:governing_equations/volumetric_flux_constraint}.
Because of the volumetric flow constraint \eqref{eq:governing_equations/volumetric_flux_constraint}, the second term of \eqref{eq:continuous_energy_conservation/j_condition} vanishes, so that the condition on the pressure gradient evaluates to
\begin{equation*}
\begin{split}
\pd{}{s} \left( Q p \right) = \pd{j}{s},
\end{split}
\end{equation*}
and \eqref{eq:continuous_energy_conservation/j_condition} is satisfied with
 \begin{equation}
\boxed{j =  Q p.}
\label{eq:continuous_energy_conservation/j_definition} 
\end{equation}

We note that $p$ is the pressure that enforces incompressibility -- it does not include a driving pressure gradient (which would appear as a source term in the TFM governing equations). 
Therefore $j$ is periodic in space in the case of periodic boundaries. 
In the case of closed boundaries, $Q$ must be zero, meaning that $j=0$ throughout the domain. 
This means that when integrating \eqref{eq:continuous_energy_conservation/continuous_energy_equation_v3} over a closed or periodic domain, the terms involving $j$ vanish, and thus this definition for $j$ is compatible with global energy conservation as described by \eqref{eq:continuous_energy_conservation/continuous_energy_equation_v4}.

The next task is to find $h$. Based on the form of $h$ for the SWE and condition  \eqref{eq:continuous_energy_conservation/h_to_f_condition}, we propose the following choice
 \begin{align}
\Aboxed{h &= g_n   q_3 \left(H - H_U \right) + g_n  q_4  H_L   + \frac{1}{2}   \frac{q_3^3}{q_1^2} + \frac{1}{2} \frac{q_4^3}{q_2^2} } \label{eq:continuous_energy_conservation/h_definition} \\
&= g_n   q_3 \left(H - H_U \right) + g_n  q_4  H_L   + \frac{1}{2}   \rho_U A_U u_U^3 + \frac{1}{2} \rho_L A_L u_L^3, \nonumber
\end{align}
which can be shown to satisfy condition \eqref{eq:continuous_energy_conservation/h_to_f_condition} by computing:
\begin{equation*}
\left[\dotp{\v{v} }{ \pd{\v{f}}{\v{q}}}\right]^T 
=
\begin{bmatrix}
-\frac{q_3^3}{q_1^3} - g_n \d{\widehat{H}_U}{A_U} \frac{q_3}{q_1} \\
-\frac{q_4^3}{q_2^3} - g_n \d{\widehat{H}_L}{A_L} \frac{q_4}{q_2} \\
\frac{3}{2} \frac{q_3^2}{q_1^2} +  g_n \d{ \widetilde{H}_U}{A_U} \\
\frac{3}{2} \frac{q_4^2}{q_2^2} + g_n  \d{ \widetilde{H}_L}{A_L}
\end{bmatrix},
\quad
\left[\pd{h}{\v{q}}\right]^T =
\begin{bmatrix}
- \frac{q_3^3}{q_1^3} -  \frac{g_n}{\rho_U} \d{H_U}{A_U} q_3 \\
-  \frac{q_4^3}{q_2^3} +  \frac{g_n}{\rho_L} \d{H_L}{A_L} q_4   \\
\frac{3}{2}  \frac{q_3^2}{q_1^2}  + g_n \left( H- H_U \right) \\
\frac{3}{2}  \frac{q_4^2}{q_2^2} + g_n H_L
\end{bmatrix}
\end{equation*}
The last two entries in these vectors are equal because of relations \eqref{eq:appendix/derivatives_of_potential_energy_intregrals_to_areas}, derived in \autoref{sec:geometric_details}. 
The first two entries are equal due to the geometric relations \eqref{eq:appendix/derivative_of_H_hat_relation_to_derivative_of_H}, which we repeat here in terms of the conserved variables $\v{q}$:
\begin{equation}
\frac{\rho_U}{q_1} \d{\widehat{H}_U }{A_U} =   \d{H_U}{A_U}, 
\qquad
\frac{\rho_L}{q_2} \d{\widehat{H}_L }{A_L} =  - \d{H_L}{A_L}.
\label{eq:continuous_energy_conservation/entropy_potential/geometric_condition_q_together}
\end{equation}
These relations follow directly from the definitions of these geometric quantities and hold for arbitrary duct geometries. 
Note that, alternatively, condition  \eqref{eq:continuous_energy_conservation/h_to_f_condition_basic} can be used (instead of \eqref{eq:continuous_energy_conservation/h_to_f_condition}), which leads to the following conditions:
\begin{equation}
\frac{\rho_U}{q_1} \pd{\widehat{H}_U }{s} =   \pd{H_U}{s}, 
\qquad
\frac{\rho_L}{q_2} \pd{\widehat{H}_L }{s} =  - \pd{H_L}{s},
\label{eq:continuous_energy_conservation/entropy_potential/geometric_condition_q_together_s}
\end{equation}
which may also be shown to be satisfied directly via application of Leibniz' rule to the definitions of $\widehat{H}_U$ and $\widehat{H}_L$.
These last two conditions will play an important role in the discrete analysis in \autoref{sec:semi-discrete_local_energy_conservation}.

In conclusion, we have proposed a novel set of $e$, $h$, and $j$ for the TFM and have shown that the local energy conservation equation \eqref{eq:continuous_energy_conservation/continuous_energy_equation_v3} is satisfied.

\subsection{Reformulation in terms of the entropy potential and conditions on fluxes}
\label{ssec:continuous_energy_conservation/entropy_potential}

Conditions \eqref{eq:continuous_energy_conservation/h_to_f_condition_basic} and \eqref{eq:continuous_energy_conservation/j_condition_basic}, or their alternatives \eqref{eq:continuous_energy_conservation/h_to_f_condition} and \eqref{eq:continuous_energy_conservation/j_condition}, were used in the previous section to find a combination of $e$, $h$ and $j$ for the continuous TFM, \textit{given} the fluxes $\v{f}$ from the governing equations. In  \autoref{sec:semi-discrete_local_energy_conservation}, we will instead aim to find discrete flux functions, \textit{given} discretized versions of $e$, $h$ and $j$ (that are inspired by their continuous counterparts). Equation \eqref{eq:continuous_energy_conservation/h_to_f_condition_basic} is not a very useful formulation to find such numerical flux functions, because it is a condition imposed on the jump in $\v{f}$, rather than $\v{f}$ itself. Therefore, equation  \eqref{eq:continuous_energy_conservation/h_to_f_condition_basic} is reformulated  using the concept of the \textit{entropy potential} \cite{FjordholmMishraTadmor2009,TadmorZhong2008}.

The entropy potential is defined to be related to $\v{v}$, $\v{f}$, and $h$ in the following manner:
\begin{equation}
\psi \coloneqq \dotp{ \v{v} }{ \v{f} } - h.
 \label{eq:continuous_energy_conservation/entropy_potential_definition}
\end{equation}
With this definition, we can reformulate condition \eqref{eq:continuous_energy_conservation/h_to_f_condition_basic} using the product rule ($\pd{}{s} \dotp{ \v{v} }{ \v{f} } = \dotp{ \pd{\v{v}}{s} }{ \v{f} } + \dotp{ \v{v} }{ \pd{\v{f}}{s} }$) as: 
\begin{equation}
\pd{\psi}{s} = \dotp{ \pd{\v{v}}{s} }{ \v{f} }.
\label{eq:continuous_energy_conservation/relation_psi_v_f_s}
\end{equation}

The entropy potential can be directly calculated from its definition \eqref{eq:continuous_energy_conservation/entropy_potential_definition} and is given by: 
\begin{align}
\Aboxed{ \psi = \dotp{\v{v}}{\v{f}} - h &= -  \rho_U g_n  \widehat{H}_U  \frac{q_3}{q_1}  - \rho_L g_n  \widehat{H}_L \frac{q_4}{q_2} } \label{eq:continuous_energy_conservation/specific_entropy_potential} \\
&= -  \rho_U g_n  \widehat{H}_U u_U  - \rho_L g_n  \widehat{H}_L u_L. \nonumber
\end{align}
Because this entropy potential is based on an $h$ that satisfies \eqref{eq:continuous_energy_conservation/h_to_f_condition_basic}, \eqref{eq:continuous_energy_conservation/relation_psi_v_f_s} is satisfied by construction. 
Nevertheless, we outline the details to convert \eqref{eq:continuous_energy_conservation/relation_psi_v_f_s} into conditions on the individual numerical fluxes, since they will be exactly mimicked by our discrete analysis in \autoref{sec:semi-discrete_local_energy_conservation}. 
We first introduce the following notation for the fluxes, and split them into the following components:
\begin{equation}
\v{f}
=
 \begin{bmatrix}
f_1(q_3) \\
f_2(q_4)  \\
  f_{3,a}(q_1,q_3) + g_n f_{3,g}(q_1) \\
 f_{4,a}(q_2,q_4) + g_n f_{4,g}(q_2)
\end{bmatrix}.
\label{eq:continuous_energy_conservation/flux_splitting}
\end{equation}
Here $f_{3,a}$ and $f_{4,a}$ are the momentum advection terms, and $f_{3,g}$ and $f_{4,g}$ are the level gradient terms (divided by $g_n$).
These fluxes and the definitions for $\v{v}$ \eqref{eq:continuous_energy_conservation/v_definition} and $\psi$ \eqref{eq:continuous_energy_conservation/specific_entropy_potential} can be substituted in \eqref{eq:continuous_energy_conservation/relation_psi_v_f_s}.
The resulting condition is first split into two conditions: one condition proportional to $g_n$, and one not proportional to $g_n$. This is done on the basis that the mass and momentum advection terms \textit{do not} depend on $g_n$ in the continuous case (see \eqref{eq:governing_equations/conservative_v1/flux_vector}), and \textit{should not} depend on $g_{n}$ in the discrete case. These two  conditions are split again on the basis that $f_1$ and $f_3$ should not depend on $q_2$ and $q_4$, and $f_2$ and $f_4$ should not depend on $q_1$ and $q_3$. We obtain the following four conditions:
\begin{subequations}
\label{eq:continuous_energy_conservation/specific_energy_1/split_condition_all_4}
\begin{align}
- \pd{}{s} \left(  \frac{1}{2}  \frac{q_3^2}{q_1^2}  \right)  f_1 + \pd{}{s} \left(  \frac{q_3}{q_1} \right) f_{3,a} &= 0, 
\label{eq:continuous_energy_conservation/specific_energy_1/split_condition_upper_no_g}  \\
- \pd{}{s} \left(  \frac{1}{2} \frac{q_4^2}{q_2^2}  \right)  f_2 + \pd{}{s} \left(\frac{q_4}{q_2} \right) f_{4,a} &= 0, 
\label{eq:continuous_energy_conservation/specific_energy_1/split_condition_lower_no_g}  \\
  \pd{}{s} \left( g_n (H-H_U) \right) f_1 + \pd{}{s} \left(  \frac{q_3}{q_1} \right) g_n f_{3,g} 
  &= - \pd{}{s} \left( \rho_U g_n  \widehat{H}_U  \frac{q_3}{q_1} \right),
  \label{eq:continuous_energy_conservation/specific_energy_1/split_condition_upper_with_g} \\
\pd{}{s} \left( g_n H_L \right) f_2 + \pd{}{s} \left(\frac{q_4}{q_2} \right) g_n f_{4,g}
&= - \pd{}{s} \left( \rho_L g_n  \widehat{H}_L \frac{q_4}{q_2} \right).
\label{eq:continuous_energy_conservation/specific_energy_1/split_condition_lower_with_g}
\end{align}
\end{subequations}

As mentioned, these equations are by construction satisfied by the flux vector \eqref{eq:governing_equations/conservative_v1/flux_vector}. One important remark is that after we reformulated in terms of $\psi$, the geometric conditions \eqref{eq:continuous_energy_conservation/entropy_potential/geometric_condition_q_together_s} encountered in \autoref{ssec:continuous_energy_conservation/specific_energy} still need to be satisfied in order for \eqref{eq:continuous_energy_conservation/specific_energy_1/split_condition_upper_with_g} and \eqref{eq:continuous_energy_conservation/specific_energy_1/split_condition_lower_with_g} to hold.

\subsection{Comparison of the energy and energy fluxes to those of other models}
\label{sec:energy_comparison}

Here we compare the  expressions obtained for $e$ and $h$ to results from literature for other models, focusing on the case of a channel geometry.
The expression \eqref{eq:continuous_energy_conservation/conservative/e_definition_1} for $e$ for the channel geometry can be obtained by substitution of the channel-specific evaluations of $\widetilde{H}_U$ and $\widetilde{H}_L$ (\autoref{sec:geometric_details}):
\begin{equation}
e_{\mathrm{ch}} =  
\rho_U g_n A_U (A - \frac{1}{2} A_U)  +  \frac{1}{2} \rho_L g_n A_L^2 + \frac{1}{2}  \rho_U  u_U^2 A_U + \frac{1}{2}\rho_L u_L^2 A_L.
\label{eq:continuous_energy_conservation/e_definition_channel_1}
\end{equation}
For a single layer fluid, such as the single layer SWE, only the third and fifth terms remain, and they are consistent with the SWE entropy function as discussed in \cite{FjordholmMishraTadmor2011} (without channel inclination).

To compare with two layer SWE theory, we rewrite \eqref{eq:continuous_energy_conservation/e_definition_channel_1} using the volume constraint \eqref{eq:governing_equations/volume_constraint} to obtain 
\begin{equation*}
e_{\mathrm{AK}} =  \frac{1}{2} \rho_U g_n A_U^2 + \rho_U g_n  A_U A_L +  \frac{1}{2} \rho_L g_n A_L^2 + \frac{1}{2}\rho_L u_L^2 A_L   + \frac{1}{2}  \rho_U  u_U^2 A_U .
\end{equation*}
This is the expression presented by Abgrall and Karni (AK) \cite{AbgrallKarni2009} and Fjordholm \cite{Fjordholm2012} as an entropy function for the two-layer SWE. The energy found in the present study can be seen as a generalization of the two-layer SWE energy to arbitrary duct geometries.

When comparing our energy flux $h$ for the TFM to the one for the two-layer SWE, it should be realized that the two-layer SWE can be obtained from the TFM by the choice $p = \rho_U g_{n} H_U$. This means that the pressure flux $j$ of the TFM needs to be added to $h$ in order to compare with the SWE expressions.
In our notation, the two-layer SWE entropy flux given by \cite{AbgrallKarni2009} is
 \begin{equation*}
\begin{split}
h_{\mathrm{AK}} &=
\rho_U g_n  u_U  A_U^2 + \rho_U g_n \left( u_U + u_L \right) A_U A_L +  \rho_L g_n u_L A_L^2 +  \frac{1}{2} \rho_U  u_U^3 A_U + \frac{1}{2} \rho_L u_L^3 A_L.
\end{split}
\end{equation*}
Our expression for $h$ for a channel is given by
\begin{equation*}
\begin{split}
h_{\mathrm{ch}} &=
\rho_U g_n u_U A_U (A - A_U) +  \rho_L g_n u_L A_L^2 +  \frac{1}{2} \rho_U  u_U^3 A_U + \frac{1}{2} \rho_L u_L^3 A_L.
\end{split}
\end{equation*}
Upon adding $j = Q p = Q \rho_U g_{n} A_U$ to $h_{\mathrm{ch}}$, and after some rewriting, we see that our TFM energy flux is consistent with the two-layer SWE entropy flux:
\begin{equation}
h_{\mathrm{AK}} = h_{\mathrm{ch}} + j.
\end{equation}

To conclude, our proposed energy \eqref{eq:continuous_energy_conservation/conservative/e_definition_1} and energy flux \eqref{eq:continuous_energy_conservation/h_definition} can be seen as a generalization of the two-layer SWE energy and energy flux to arbitrary duct cross sections. 

\section{Discretization of the governing equations}
\label{sec:discretization}

\subsection{Semi-discrete model equations}
\label{ssec:numerical_scheme/semi-discrete-model-staggered}

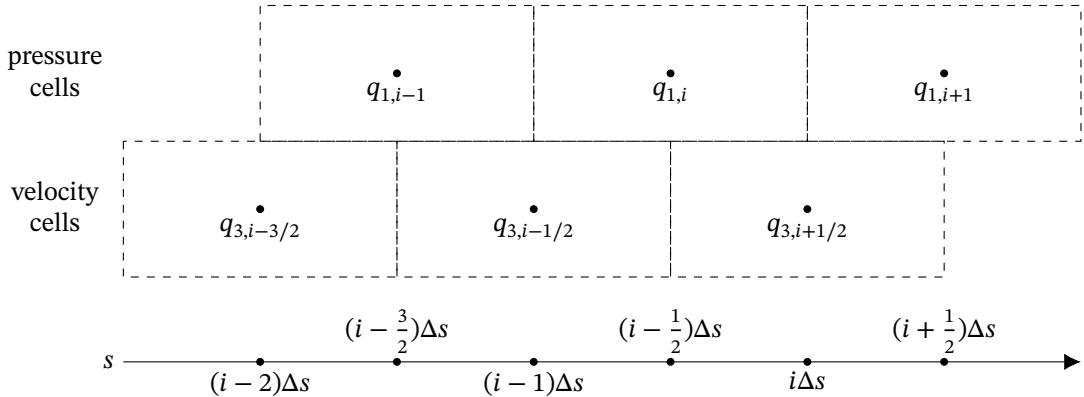
\begin{figure}[hbtp]
\centering%
\begin{tikzpicture}[scale=0.9]

\node at (-11,1.25) {\begin{tabular}{c} pressure \\ cells \end{tabular}};
\node at (-11,-0.75) {\begin{tabular}{c} velocity \\ cells \end{tabular}};

\draw[dashed] (-8,0.25) rectangle (-4,2.25);
\draw[fill=black] (-6,1.25) circle (0.05); 
\node [below] at (-6,1.25) {$q_{1,i-1}$};

\draw[dashed] (-4,0.25) rectangle (-0,2.25);
\draw[fill=black] (-2,1.25) circle (0.05); 
\node [below] at (-2,1.25) {$q_{1,i}$};

\draw[dashed] (-0,0.25) rectangle (4,2.25);
\draw[fill=black] (2,1.25) circle (0.05); 
\node [below] at (2,1.25) {$q_{1,i+1}$};


%
%

\draw[dashed] (-10,0.25) rectangle (-6,-1.75);
\draw[fill=black] (-8,-0.75) circle (0.05); 
\node [below] at (-8,-0.75) {$q_{3,i-3/2}$};

\draw[dashed] (-6,0.25) rectangle (-2,-1.75);
\draw[fill=black] (-4,-0.75) circle (0.05); 
\node [below] at (-4,-0.75) {$q_{3,i-1/2}$};

\draw[dashed] (-2,0.25) rectangle (2,-1.75);
\draw[fill=black] (-0,-0.75) circle (0.05); 
\node [below] at (0,-0.75) {$q_{3,i+1/2}$};


\node [left] at (-10,-3) {$s$};
\draw[-{Latex[width = 2.2mm, length = 2.2mm]}] (-10, -3)--(4,-3);

\draw[fill=black] (-8,-3) circle (0.05); 
\node [below] at (-8,-3) {$(i-2)\Delta s$};

\draw[fill=black] (-6,-3) circle (0.05); 
\node [above] at (-6,-3) {$(i-\frac{3}{2})\Delta s$};

\draw[fill=black] (-4,-3) circle (0.05); 
\node [below] at (-4,-3) {$(i-1)\Delta s$};

\draw[fill=black] (-2,-3) circle (0.05); 
\node [above] at (-2,-3) {$(i-\frac{1}{2})\Delta s$};

\draw[fill=black] (0,-3) circle (0.05); 
\node [below] at (0,-3) {$i\Delta s$};

\draw[fill=black] (2,-3) circle (0.05); 
\node [above] at (2,-3) {$(i+\frac{1}{2})\Delta s$};


\normalsize
\end{tikzpicture}
\caption{Staggered grid layout.\label{fig:staggered_grid}}
\end{figure}

The system of equations \eqref{eq:governing_equations/conservative_v1} is discretized using a finite volume method on a uniform staggered grid, sketched in \autoref{fig:staggered_grid}.
This discretization naturally conserves mass for each fluid separately, and momentum for both fluids combined. 
The first two components of $\v{q}$ (the phase masses) and the pressure are defined at the centers of $N_p$ pressure volumes, which have a cell size of $\Delta s = L/N_p$. The last two components of $\v{q}$ (the phase momenta) are defined at the centers of $N_u$ velocity volumes. 

%

On this staggered grid we define a local discrete vector of unknowns as follows:
\begin{equation}
\v{q}_i(t) 
\coloneqq
\begin{bmatrix}
q_{1,i}(t) \\
q_{2,i}(t) \\
q_{3,i-1/2}(t) \\
q_{4,i-1/2}(t) 
\end{bmatrix}
=
\begin{bmatrix}
\left(\rho_U A_U \Delta s \right)_i \\
\left(\rho_L A_L \Delta s \right)_i \\
\left(\rho_U A_U u_U \Delta s \right)_{i-1/2} \\
\left(\rho_L A_L u_L \Delta s \right)_{i-1/2}
\end{bmatrix}
=
\begin{bmatrix}
\rho_U A_{U,i} \Delta s \\
\rho_L A_{L,i} \Delta s  \\
\rho_U \overline{A}_{U,i-1/2} u_{U,i-1/2} \Delta s \\
\rho_L \overline{A}_{L,i-1/2} u_{L,i-1/2} \Delta s 
\end{bmatrix}.
\label{eq:discretization/discrete_variables_u/local}
\end{equation}
The choice of using $i-1/2$ in the definition of $\v{q}_i$ instead of $i+1/2$ is arbitrary. 
Note that $q_{1,i}(t)\approx q_{1}(s_{i},t)$ (and similar for the other entries); the notation is on purpose kept very close to the notation of the continuous model, but can be distinguished due to the extra index which the discrete variables carry. 
Another notable difference is that the cell sizes are included in the discrete unknowns, so that they have units of mass and momentum. 

The last equality in \eqref{eq:discretization/discrete_variables_u/local} describes the relations of the discrete conservative variables to the discrete primitive variables (cross-sections and velocities). 
Here we have introduced the following notation for interpolation operators \cite{FjordholmMishraTadmor2009}:
\begin{equation}
\overline{a}_{i-1/2} \coloneqq \frac{1}{2} \left( a_{i-1} + a_{i} \right) \quad \overline{a}_{i} \coloneqq \frac{1}{2} \left( a_{i-1/2} + a_{i+1/2} \right). 
\end{equation} 
The numerical scheme is implemented in terms of the conservative variables $q_{1,i}$ through $q_{4,i-1/2}$, but the primitive variables  can be extracted in post-processing according to the given relations.

The notation with $\v{q}_i$ as a discrete local vector of unknowns allows us to write the discrete scheme in vector form as 
 \begin{equation}
\d{\v{q}_{i}}{t} + \left( \v{f}_{i+1/2} - \v{f}_{i-1/2} \right) + \v{d}_i \left( p_i - p_{i-1} \right)= \v{0}. 
 \label{eq:discretization/staggered/finite_volume_scheme}
\end{equation}
Here, we have defined $\v{f}_{i-1/2}$ as 
\begin{equation*}
\v{f}_{i-1/2}(\v{q}_{i-2},\v{q}_{i-1},\v{q}_i) \coloneqq 
\begin{bmatrix}
f_{1,i-1/2}(\v{q}_i) \\
f_{2,i-1/2}(\v{q}_i) \\
f_{3,i-1}(\v{q}_{i-2},\v{q}_{i-1},\v{q}_i) \\
f_{4,i-1}(\v{q}_{i-2},\v{q}_{i-1},\v{q}_i)
\end{bmatrix},
\end{equation*}
and 
\begin{equation*}
\v{d}_{i}(\v{q}_{i-1},\v{q}_i) \coloneqq 
\begin{bmatrix}
0 \\
0\\
d_{3,i-1/2}(\v{q}_{i-1},\v{q}_i) \\
d_{4,i-1/2}(\v{q}_{i-1},\v{q}_i)
\end{bmatrix}.
\end{equation*}
The numerical fluxes and numerical pressure terms are left undefined in this section, because we will define them based on the requirement of energy conservation, in \autoref{sec:semi-discrete_local_energy_conservation}. 

The pressure terms are non-conservative and are not written as the difference between an inflow and an outflow of the finite volume cell.
However, with the staggered grid employed here, one can see that $q_{3,i-1/2}$ and $q_{4,i-1/2}$ are directly and naturally connected to the pressure at the neighboring grid cells. Analogous to the incompressible (multi-dimensional) single-phase Navier-Stokes equations (for which staggered grids are known to lead to strong coupling), this pressure-velocity coupling is necessary to prevent checkerboard patterns, and would be much more difficult to achieve on a collocated grid.


The system is closed by the volume constraint (compare to \eqref{eq:governing_equations/volume_constraint}):
\begin{equation}
\frac{q_{1,i}}{\rho_U \Delta s} + \frac{q_{2,i}}{\rho_L \Delta s}  = A,
\label{eq:discrete_volume_constraint}
\end{equation}
which implies the volumetric flow constraint (compare to \eqref{eq:governing_equations/volumetric_flux_constraint_diff})
\begin{equation}
Q_{i+1/2} - Q_{i-1/2} \coloneqq\frac{q_{3,i+1/2}}{\rho_U \Delta s} - \frac{q_{3,i-1/2}}{\rho_U \Delta s} + \frac{q_{4,i+1/2}}{\rho_L \Delta s} -  \frac{q_{4,i-1/2}}{\rho_L \Delta s} = 0,
\label{eq:discrete_volumetric_flow_constraint}
\end{equation}
so that $Q_{i+1/2} = Q_{i-1/2} = Q(t)$, like in the continuous case.
This step can only be made if we choose $f_{1,i-1/2} = q_{3,i-1/2}/\Delta s$ and $f_{2,i-1/2} = q_{4,i-1/2}/\Delta s$, and this will be used as a condition on the form of the numerical fluxes in \autoref{ssec:semi-discrete_local_energy_conservation/derivation_of_numerical_fluxes}. 

Just as in the continuous case, these constraints are used to set up a Poisson equation for the pressure. 
The semi-discrete momentum equations are first summed to obtain
\begin{equation}
\frac{1}{\Delta s} \v{l}^T \v{d}_i  \left( p_i - p_{i-1} \right) = - \frac{1}{\Delta s}\v{l}^T \left( \d{\v{q}_i}{t}  +  \left( \v{f}_{i+1/2} - \v{f}_{i-1/2} \right) \right), \quad \text{with} \quad 
\v{l}^T = 
\begin{bmatrix}
0 & 0 & \frac{1}{\rho_U } & \frac{1}{\rho_L }
\end{bmatrix}.
  \label{eq:discrete_pressure_Poisson_pre_pre_step}
\end{equation}
Expanding and substituting the definition of $Q_{i-1/2}$ yields
 \begin{equation}
 \frac{1}{\Delta s} \left( \frac{d_{3,i-1/2}}{\rho_U} + \frac{d_{4,i-1/2}}{\rho_L} \right)\left( p_i - p_{i-1} \right) = -\d{Q_{i-1/2}}{t}  -  \frac{1}{\Delta s} \left( \frac{f_{3,i}-f_{3,i-1}}{\rho_U } + \frac{f_{4,i}-f_{4,i-1}}{\rho_L } \right).
 \label{eq:discrete_pressure_Poisson_pre_step}
 \end{equation}
After taking the difference between this equation and the same equation for index $i+1/2$, and applying \eqref{eq:discrete_volumetric_flow_constraint}, we obtain the discrete version of \eqref{eq:governing_equations/pressure_Poisson}:
  \begin{multline}
 \frac{1}{\Delta s^2} \left[\left( \frac{d_{3,i+1/2}}{\rho_U} + \frac{d_{4,i+1/2}}{\rho_L} \right)\left( p_{i+1} - p_{i} \right) - \left( \frac{d_{3,i-1/2}}{\rho_U} + \frac{d_{4,i-1/2}}{\rho_L} \right)\left( p_i - p_{i-1} \right) \right] \\
 =  -  \frac{1}{\Delta s^2} \left( \frac{f_{3,i+1}-2f_{3,i}+f_{3,i-1}}{\rho_U } + \frac{f_{4,i+1}-2f_{4,i}+f_{4,i-1}}{\rho_L} \right).
 \label{eq:discrete_pressure_Poisson}
 \end{multline}

System \eqref{eq:discretization/staggered/finite_volume_scheme} is discretized in time using the fourth-order semi-explicit Runge-Kutta method described in \cite{SanderseVeldman2019}.
At each stage of the Runge-Kutta time step, a predictor-corrector algorithm is applied: the momentum equations are first solved without including the pressure terms, the discrete Poisson equation is solved for the pressure using these intermediate momenta, and the momenta are updated in a projection step using the calculated pressure.
This ensures that the volume and volumetric flow constraints are satisfied at all stages.  
We solve \eqref{eq:discrete_pressure_Poisson} iteratively, using a preconditioned conjugate gradient method. 
The time integration method is fourth-order accurate for all variables, and requires a restriction to the CFL-number based on the eigenvalues of the TFM.

\subsection{Boundary conditions}
In the case of periodic boundaries, the domain is divided into $N_p$ pressure volumes and $N_u=N_p$ velocity volumes. 
There are no special boundary points: the scheme as laid out in \autoref{ssec:numerical_scheme/semi-discrete-model-staggered} applies everywhere, looping around the domain. 

For closed boundaries, there are $N_p$ interior pressure points and $N_u=N_p-1$ interior velocity points. 
The first interior pressure node is located at $s=\Delta s/2$, the first interior velocity node is located at $s=\Delta s$, and similarly for the last nodes at the end of the domain \cite{SanderseVeldman2019}. 
For both the pressure and velocity grids, there are boundary points in addition to the interior points, one at each side of the domain. 
When calculating the discrete energy on the velocity grid (see \autoref{sec:semi-discrete_local_energy_conservation/outline}), it is important to include the half-volumes between the boundary points and the first and last interior points. 

At the boundary points, the mass fluxes ($\rho_U A_U u_U$ and $\rho_L A_L u_L$) are specified, and $A_U$ and $A_L$ follow via an analysis of the characteristics corresponding to the incoming and outgoing waves at the boundary. In the case of closed boundaries, as used in this work, the mass fluxes are set to zero.
Note that the characteristic analysis incorporates the volume constraint \eqref{eq:discrete_volume_constraint}, and no boundary condition is needed for the pressure (the pressure at the boundaries has no influence on the solution in the interior). For more details on the implementation of the boundary conditions we refer to \cite{SanderseVeldman2019}. 

\section{Energy-conserving spatial discretization of the two-fluid model}
\label{sec:semi-discrete_local_energy_conservation}

\subsection{Outline: conditions for discrete energy conservation}
\label{sec:semi-discrete_local_energy_conservation/outline}

In the discrete case, just as in the continuous case, we want to satisfy a local and global energy equality. The use of a staggered grid instead of the commonly used collocated grid (\eg \cite{FjordholmMishraTadmor2009,TadmorZhong2008}) makes it straightforward to obtain an energy-conserving discretization of the non-conservative pressure term, but introduces new challenges in terms of the definition of the discrete local energy, which is not unique anymore.

We choose to define the local energy at the velocity grid points, \ie we choose $e_{i-1/2} = e(\v{q}_{i-1},\v{q}_{i})$, and are aiming for a discrete version of \eqref{eq:continuous_energy_conservation/continuous_energy_equation_v3}:
\begin{equation}
\d{e_{i-1/2}}{t} + \left( h_{i} - h_{i-1} \right) + \left( j_{i} - j_{i-1} \right)= 0,
\label{eq:semi-discrete_local_energy_conservation/local_energy_conservation_equation}
\end{equation}
with $h_{i} = h(\v{q}_{i-1},\v{q}_{i})$ and $j_{i} = j(p_i)$ as the numerical energy fluxes. 
This choice means that the potential energy terms and $q_1$ and $q_2$ in the kinetic energy terms need to be interpolated, but $q_3$ and $q_4$ do not require interpolation. 
With this choice, we obtain energy-conserving expressions for $f_{3,i}$ and $f_{4,i}$ in a constructive manner (after choosing advantageous expressions for $f_{1,i-1/2}$ and $f_{2,i-1/2}$). 
It is also possible to define the energy at the pressure grid points, and obtain an energy-conserving discretization, but in that case it is necessary to substitute trial solutions for $f_{3,i}$ and $f_{4,i}$, and interpolation of the pressure is required in the expression for $j$ (see the remark at the end of \autoref{ssec:semi-discrete_local_energy_conservation/derivation_of_numerical_fluxes}).
We would like to emphasize that \eqref{eq:semi-discrete_local_energy_conservation/local_energy_conservation_equation} is \textit{not} being solved as an additional equation; instead it will be shown to be \textit{a consequence} of the discrete mass and momentum equations given in \autoref{sec:discretization}, if the numerical fluxes and $\v{d}_i$ are chosen appropriately.

If \eqref{eq:semi-discrete_local_energy_conservation/local_energy_conservation_equation} holds, it can be summed over all finite volumes to yield
\begin{align}
\d{E_{h}}{t} = \sum_{i=1}^{N_u} \d{e_{i-1/2}}{t} &= - \sum_{i=1}^{N_u} \left[ \left( h + j\right)_i - \left( h + j\right)_{i-1} \right] = 0, 
\label{eq:semi-discrete_local_energy_conservation/global_energy_conservation_equation} 
\end{align}
where the last equality should hold in the case of periodic or closed boundaries. Here we have defined the global discrete energy as the discrete counterpart of \eqref{eq:continuous_energy_conservation/definition}:
\begin{equation}
E_{h} = E_{h}(t) \coloneqq \sum_{i=1}^{N_u} e_{i-1/2}.
\label{eq:semi-discrete_local_energy_conservation/global_energy_definition}
\end{equation}

Like in the continuous case, the art is to find expressions for $e_{i-1/2}$, $h_{i}$ and $j_{i}$ such that equation \eqref{eq:semi-discrete_local_energy_conservation/local_energy_conservation_equation} is satisfied. In addition, the numerical flux $\v{f}_{i-1/2}$ needs to be constructed. We will outline the steps to obtain these quantities in a manner parallel to the continuous derivation in \autoref{sec:continuous_energy_conservation}.

First, we postulate an energy 
\begin{equation}
e_{i-1/2} =e_{i-1/2}\left(\v{q}_{i-1}, \v{q}_i \right),
\label{eq:semi-discrete_local_energy_conservation/general_staggered_shifted_energy}
\end{equation}
which will be based on the energy found for the continuous case. Note that the dependence could be expanded to additional grid points if required, but we will introduce an energy for which this is not necessary.
Second, calculate the vectors of entropy variables, defined as
\begin{equation*}
\v{v}_{i-1/2,i} \coloneqq \left[\pd{e_{i-1/2}}{\v{q}_i}\right]^T, \quad
\v{v}_{i-1/2,i-1} \coloneqq \left[\pd{e_{i-1/2}}{\v{q}_{i-1}}\right]^T.
\end{equation*}
Here the first index refers to the index of the energy, and the second index refers to the conservative variables to which derivatives are taken.

 For the energy given by \eqref{eq:semi-discrete_local_energy_conservation/general_staggered_shifted_energy}, 
the time derivative can be expressed as 
\begin{equation}
 \d{e_{i-1/2}}{t} =  \dotp{ \v{v}_{i-1/2,i-1} }{ \d{\v{q}_{i-1} }{t} }  + \dotp{ \v{v}_{i-1/2,i} }{\d{\v{q}_i}{t} },
 \label{eq:semi-discrete_local_energy_conservation/de_dt}
\end{equation}
 Here the brackets represent dot products over the vectors (at a certain grid point), just as in the continuous case.
The right-hand side of equation \eqref{eq:semi-discrete_local_energy_conservation/de_dt} follows by substituting equation \eqref{eq:discretization/staggered/finite_volume_scheme} for $i$ and $i-1$:
\begin{equation}
 \d{e_{i-1/2}}{t} =
 \left< \v{v}_{i-1/2,i-1}, \llbracket \v{f}_{i-1} \rrbracket \right>  + \dotp{ \v{v}_{i-1/2,i-1} }{\v{d}_{i-1} } \llbracket p_{i-3/2}  \rrbracket + \left< \v{v}_{i-1/2,i}, \llbracket \v{f}_{i} \rrbracket \right> + \dotp{\v{v}_{i-1/2,i} }{ \v{d}_i } \llbracket p_{i-1/2} \rrbracket,
  \label{eq:semi-discrete_local_energy_conservation/energy_equation_v1}
 \end{equation}
where we have introduced the following notation for jump operators \cite{FjordholmMishraTadmor2009}:
\begin{equation}
\llbracket a_{i-1/2} \rrbracket  \coloneqq a_{i} - a_{i-1}, \quad \llbracket a_{i} \rrbracket  \coloneqq a_{i+1/2} - a_{i-1/2}.
\end{equation} 
Comparing with \eqref{eq:semi-discrete_local_energy_conservation/local_energy_conservation_equation}
 we see that the energy fluxes $h_i$ and $j_i$ need to satisfy
\begin{gather}
\left< \v{v}_{i-1/2,i-1}, \llbracket \v{f}_{i-1} \rrbracket \right>  +  \left< \v{v}_{i-1/2,i}, \llbracket \v{f}_{i} \rrbracket \right> = \left\llbracket h_{i-1/2} \right\rrbracket,
\label{eq:semi-discrete_local_energy_conservation/h_to_f_condition_basic} \\
\dotp{ \v{v}_{i-1/2,i-1} }{\v{d}_{i-1} } \llbracket p_{i-3/2}  \rrbracket + \dotp{\v{v}_{i-1/2,i} }{ \v{d}_i } \llbracket p_{i-1/2} \rrbracket = \llbracket j_{i-1/2} \rrbracket.
\label{eq:semi-discrete_local_energy_conservation/j_condition_basic}
\end{gather}
These conditions are analogous to \eqref{eq:continuous_energy_conservation/h_to_f_condition_basic} and  \eqref{eq:continuous_energy_conservation/j_condition_basic} for the continuous case, with discrete jumps corresponding to derivatives with respect to $s$. 
Together, conditions \eqref{eq:semi-discrete_local_energy_conservation/h_to_f_condition_basic} and \eqref{eq:semi-discrete_local_energy_conservation/j_condition_basic} guarantee that  \eqref{eq:semi-discrete_local_energy_conservation/energy_equation_v1} can be written as \eqref{eq:semi-discrete_local_energy_conservation/local_energy_conservation_equation},
thus proving conservation of the discrete local energy \eqref{eq:semi-discrete_local_energy_conservation/general_staggered_shifted_energy}.

The challenge is to find the proper combination of discrete expressions for $e_{i-1/2}$, $h_i$, $j_i$, and $\v{f}_{i-1/2}$ which are consistent approximations to their continuous counterparts in such a way that the local energy conservation equation is satisfied. 
This is a difficult problem, since we have multiple degrees of freedom ($e_{i-1/2}$, $h_i$, $j_i$, and $\v{f}_{i-1/2}$), and the solution might not be unique. To simplify the construction, we will use the concept of entropy potential introduced in \autoref{ssec:continuous_energy_conservation/entropy_potential}: after choosing a certain $e_{i-1/2}$ and $\psi_{i-1/2}$, this yields straightforward conditions on the fluxes $\v{f}_{i-1/2}$ to be energy-conserving.

\subsection{Choice of discrete energy and energy fluxes}
\label{sec:semi-discrete_local_energy_conservation/choice}

In this section we propose an energy $e_{i-1/2}$, and verify that this energy is conserved by the pressure terms of the discrete model (energy conservation for the flux terms is treated in \autoref{ssec:semi-discrete_local_energy_conservation/entropy_potential} and \autoref{ssec:semi-discrete_local_energy_conservation/derivation_of_numerical_fluxes}).
Recalling the continuous energy \eqref{eq:continuous_energy_conservation/conservative/e_definition_1}, we define a discrete energy  $e_{i-1/2} = e_{i-1/2}(\v{q}_{i-1},\v{q}_i)$:
\begin{align}
\Aboxed{
e_{i-1/2}  &= \rho_U g_n \overline{\widetilde{H}}_{U,i-1/2} \Delta s +  \rho_L g_n \overline{\widetilde{H}}_{L,i-1/2} \Delta s + \frac{1}{2} \frac{q_{3,i-1/2}^2}{ \overline{q}_{1,i-1/2}  } + \frac{1}{2} \frac{q_{4,i-1/2}^2}{  \overline{q}_{2,i-1/2}  }  }
\label{eq:semi-discrete_local_energy_conservation/specific_energy_1/energy_definition} 
\\ &= \rho_U g_n \overline{\widetilde{H}}_{U,i-1/2} \Delta s +  \rho_L g_n \overline{\widetilde{H}}_{L,i-1/2} \Delta s + \frac{1}{2} \rho_U \overline{A}_{U,i-1/2} u_{U,i-1/2}^2 \Delta s + \frac{1}{2} \rho_L \overline{A}_{L,i-1/2} u_{L,i-1/2}^2 \Delta s. \nonumber
\end{align}
Other choices are possible because on a staggered grid interpolation is required, and the interpolation may be carried out in various different ways\footnote{In \autoref{sec:global_discrete_energy_conservation} we will show that the same results can be obtained with a global energy analysis, in which interpolation of the local potential energy to the velocity grid points is not needed.}. 
Our choice \eqref{eq:semi-discrete_local_energy_conservation/specific_energy_1/energy_definition} is one of the most straightforward choices for the energy that is consistent with the continuous definition, when the energy is defined at the velocity grid points, and leads to an elegant form of the energy-conserving discretization (see also the remark at the end of \autoref{ssec:semi-discrete_local_energy_conservation/derivation_of_numerical_fluxes}).

We use identities given in \autoref{sec:geometric_details} to calculate the $\v{v}$ vectors.
They are given by
\begin{equation}
\v{v}_{i-1/2,i-1} = \left[\pd{e_{i-1/2}}{\v{q}_{i-1}}\right]^T =
\begin{bmatrix}
- \frac{1}{4 } \frac{q_{3,i-1/2}^2}{  \overline{q}_{1,i-1/2}^2 } + \frac{1}{2} g_n \left(H-H_{U,i-1} \right) \\
- \frac{1}{4} \frac{q_{4,i-1/2}^2}{  \overline{q}_{2,i-1/2}^2 } + \frac{1}{2} g_n H_{L,i-1} \\
0\\
0
\end{bmatrix},
\label{eq:semi-discrete_local_energy_conservation/specific_energy_1/v_i_min_1}
\end{equation}
and
\begin{equation}
\v{v}_{i-1/2,i} = \left[\pd{e_{i-1/2}}{\v{q}_i}\right]^T =
\begin{bmatrix}
- \frac{1}{4 } \frac{q_{3,i-1/2}^2}{  \overline{q}_{1,i-1/2}^2 } + \frac{1}{2 } g_n \left(H-H_{U,i}\right) \\
- \frac{1}{4 } \frac{q_{4,i-1/2}^2}{  \overline{q}_{2,i-1/2}^2 } + \frac{1}{2} g_n H_{L,i} \\
\frac{q_{3,i-1/2}}{\overline{q}_{1,i-1/2}} \\
 \frac{q_{4,i-1/2}}{\overline{q}_{2,i-1/2}} 
\end{bmatrix},
\label{eq:semi-discrete_local_energy_conservation/specific_energy_1/v_i}
\end{equation}
and their sum is consistent with \eqref{eq:continuous_energy_conservation/v_definition}.

The pressure terms in \eqref{eq:semi-discrete_local_energy_conservation/energy_equation_v1} need to satisfy condition \eqref{eq:semi-discrete_local_energy_conservation/j_condition_basic}, which can be rewritten to obtain the discrete version of \eqref{eq:continuous_energy_conservation/j_condition}:
\begin{multline}
  \left\llbracket j_{i-1/2} \right\rrbracket = 
 \llbracket  \overline{ \dotp{ \v{v}_{i-1/2,i-1} }{ \v{d}_{i-1}  } } p_{i-3/2}  \rrbracket -  \overline{ \left( \llbracket \dotp{ \v{v}_{i-1/2,i-1} }{\v{d}_{i-1} } \rrbracket p_{i-3/2} \right) } \\
 + \llbracket \overline{   \dotp{ \v{v}_{i-1/2,i} }{ \v{d}_{i}  } } p_{i-1/2}  \rrbracket -  \overline{ \left( \llbracket \dotp{ \v{v}_{i-1/2,i} }{ \v{d}_{i} } \rrbracket p_{i-1/2} \right) }.
 \label{eq:semi-discrete_local_energy_conservation/j_condition}
 \end{multline}
On a staggered grid, it is straightforward to satisfy this condition by choosing for $\v{d}_i$
\begin{equation}
\v{d}_i =
\frac{1}{\Delta s}
\begin{bmatrix}
0 \\
0\\
 \frac{\overline{q}_{1,i-1/2} }{\rho_U} \\
 \frac{\overline{q}_{2,i-1/2} }{\rho_L } 
\end{bmatrix}
=
\begin{bmatrix}
0 \\
0 \\
\overline{A}_{U,i-1/2}  \\
\overline{A}_{L,i-1/2} 
\end{bmatrix},
\label{eq:semi-discrete_local_energy_conservation/specific_energy_1/pressure_d_discretization}
\end{equation}
since with this choice we have
\begin{equation*}
\left< \v{v}_{i-1/2,i-1}, \v{d}_{i-1} \right> = 0, 
\end{equation*}
and
\begin{equation*}
\dotp{ \v{v}_{i-1/2,i}}{ \v{d}_i } = 
 \left( \frac{q_{3,i-1/2}}{\rho_U \Delta s} + \frac{q_{4,i-1/2}}{\rho_L \Delta s}  \right)=  Q_{i-1/2}.
\end{equation*}
Consequently, condition \eqref{eq:semi-discrete_local_energy_conservation/j_condition} can be written with the volumetric flow constraint \eqref{eq:discrete_volumetric_flow_constraint} as
\begin{equation}
    \llbracket j_{i-1/2} \rrbracket = \llbracket \overline{Q}_{i-1/2} p_{i-1/2} \rrbracket = \llbracket Q(t) p_{i-1/2} \rrbracket,
\end{equation}
so that \eqref{eq:semi-discrete_local_energy_conservation/j_condition} (and  \eqref{eq:semi-discrete_local_energy_conservation/j_condition_basic}) is satisfied when $j$ is given by
\begin{equation}
\boxed{j_i = Q(t) p_{i}.}
\end{equation}
Note that our constraint-consistent time integration method enforces that the volumetric flow constraint is satisfied up to machine precision \cite{SanderseVeldman2019}.


\subsection{Reformulation in terms of the entropy potential and conditions on numerical fluxes}
\label{ssec:semi-discrete_local_energy_conservation/entropy_potential}

The objective of finding energy-conserving numerical fluxes is better served by reformulating condition \eqref{eq:semi-discrete_local_energy_conservation/h_to_f_condition_basic} in terms of the entropy potential, because this results in an alternative, constructive, condition for finding energy-conserving fluxes.
The fluxes are then based on the entropy potential $\psi_{i-1/2}$ instead of the energy flux $h_i$. Similar to  \autoref{ssec:continuous_energy_conservation/entropy_potential}, we rewrite the left-hand side of  \eqref{eq:semi-discrete_local_energy_conservation/h_to_f_condition_basic} as:
\begin{equation}
\begin{split}
 \left< \v{v}_{i-1/2,i-1}, \llbracket \v{f}_{i-1} \rrbracket \right>  +  \left< \v{v}_{i-1/2,i}, \llbracket \v{f}_{i} \rrbracket \right> = 
 & \left\llbracket \dotp{ \overline{\v{v}}_{i-1/2,i-1} }{ \v{f}_{i-1} } \right\rrbracket  +  \left\llbracket \dotp{ \overline{\v{v}}_{i-1/2,i} }{ \v{f}_{i} }  \right\rrbracket \\
 & - \overline{ \dotp{\llbracket \v{v}_{i-1/2,i-1} \rrbracket }{ \v{f}_{i-1} }} - \overline{  \dotp{ \llbracket \v{v}_{i-1/2,i} \rrbracket }{ \v{f}_{i} } },
 \label{eq:semi-discrete_local_energy_conservation/energy_conservation_proof/part_1}
 \end{split}
\end{equation}
which can be interpreted as a discrete version of the product rule $\dotp{\v{v}}{\pd{\v{f}}{s}  } = \pd{}{s} \dotp{ \v{v} }{ \v{f} } - \dotp{ \pd{\v{v}}{s} }{ \v{f} }$. We have made use of the following definitions:
\begin{align}
\overline{\v{v}}_{i,i-1/2} &= \frac{1}{2} \left( \v{v}_{i-1/2,i-1} + \v{v}_{i+1/2,i} \right), \quad &\quad \overline{\v{v}}_{i,i+1/2} &= \frac{1}{2} \left( \v{v}_{i-1/2,i} + \v{v}_{i+1/2,i+1} \right), \label{eq:semi-discrete_local_energy_conservation/v_interpolation_definition}  \\
\llbracket \v{v}_{i,i-1/2} \rrbracket &= \v{v}_{i+1/2,i} - \v{v}_{i-1/2,i-1}, \quad &\quad \llbracket \v{v}_{i,i+1/2} \rrbracket &= \v{v}_{i+1/2,i+1} - \v{v}_{i-1/2,i}.
\label{eq:semi-discrete_local_energy_conservation/v_jump_definition} 
\end{align}
These definitions are such that we only interpolate or take jumps between $\v{v}$ vectors with the same relative indices. 

Instead of directly choosing $\psi$, it is more natural to use the last terms in \eqref{eq:semi-discrete_local_energy_conservation/energy_conservation_proof/part_1} and define the jump in $\psi$ (similar to \eqref{eq:continuous_energy_conservation/relation_psi_v_f_s}) as
\begin{equation}
 \llbracket \psi_{i} \rrbracket  = \dotp{ \llbracket  \v{v}_{i,i-1/2} \rrbracket }{ \v{f}_{i-1/2} } + \dotp{ \llbracket  \v{v}_{i,i+1/2} \rrbracket }{ \v{f}_{i+1/2} }, 
\label{eq:semi-discrete_local_energy_conservation/conservation_condition/2}
\end{equation}
since this leads to the following `implied' definition of $\psi$:
\begin{equation}
\llbracket \overline{\psi}_{i-1/2} \rrbracket = 
\overline{\llbracket \psi_{i-1/2} \rrbracket} = 
\frac{1}{2} \llbracket \psi_{i-1} \rrbracket + \frac{1}{2 }\llbracket \psi_{i} \rrbracket  =
\left\llbracket\dotp{\overline{\v{v}}_{i-1/2,i-1} }{\v{f}_{i-1} } \right\rrbracket + \left\llbracket \dotp{ \overline{\v{v}}_{i-1/2,i}}{ \v{f}_{i} } \right\rrbracket  -  \left\llbracket h_{i-1/2} \right\rrbracket,
\label{eq:semi-discrete_local_energy_conservation/psi_jump_definition}
\end{equation} 
which is consistent with \eqref{eq:continuous_energy_conservation/entropy_potential_definition}. The advantage of \eqref{eq:semi-discrete_local_energy_conservation/conservation_condition/2} over \eqref{eq:semi-discrete_local_energy_conservation/h_to_f_condition_basic} is that we have a condition on the flux itself, rather than on the jump in the flux. Once $e_{i-1/2}$ and $\psi_{i-1/2}$ have been chosen and $\v{f}_{i-1/2}$ has been derived, $h_{i}$ can be determined from 
\begin{equation}
h_i = \dotp{ \overline{\v{v}}_{i,i-1/2} }{ \v{f}_{i-1/2} }  + \dotp{ \overline{\v{v}}_{i,i+1/2} }{ \v{f}_{i+1/2} }  - \overline{\psi}_{i}.
\label{eq:semi-discrete_local_energy_conservation/psi_split_definition}
\end{equation}
We note that this expression is similar to the collocated grid setting, where one has \linebreak $h_{i-1/2} = \dotp{\overline{\v{v}}_{i-1/2}}{\v{f}_{i-1/2}} - \overline{\psi}_{i-1/2}$ \cite{FjordholmMishraTadmor2009}. The difference lies in a shift in indices (because our energy is defined at $i-1/2$ instead of $i$), and in the way the term $\dotp{\v{v}}{\v{f}}$ is approximated.

We propose now the following discrete entropy potential for the equations:
\begin{align}
\Aboxed{ \psi_{i-1/2}(\v{q}_{i-1}, \v{q}_{i})   &= -  \rho_U g_n  \overline{\widehat{H}}_{U,i-1/2}  \frac{q_{3,i-1/2}}{\overline{q}_{1,i-1/2}}  - \rho_L g_n  \overline{\widehat{H}}_{L,i-1/2} \frac{q_{4,i-1/2}}{\overline{q}_{2,i-1/2}} } 
 \label{eq:semi-discrete_local_energy_conservation/first_specific_energy/psi_definition_direct} \\
 &= -  \rho_U g_n  \overline{\widehat{H}}_{U,i-1/2}  u_{U,i-1/2}  - \rho_L g_n  \overline{\widehat{H}}_{L,i-1/2} u_{L,i-1/2}. \nonumber
\end{align}
This is a straightforward discretization of \eqref{eq:continuous_energy_conservation/specific_entropy_potential}. Given the expressions for $\v{v}$ (\eqref{eq:semi-discrete_local_energy_conservation/specific_energy_1/v_i_min_1} and \eqref{eq:semi-discrete_local_energy_conservation/specific_energy_1/v_i}), condition  \eqref{eq:semi-discrete_local_energy_conservation/conservation_condition/2} can now be evaluated to yield the numerical fluxes $\v{f}$. In order to be able to derive from the (scalar) condition \eqref{eq:semi-discrete_local_energy_conservation/conservation_condition/2} multiple equations for the individual numerical fluxes, we split $f_{3,i-1}$ and $f_{4,i-1}$ into an advective component (denoted by subscript $a$) and a level gradient (or gravity) component (denoted by subscript $g$): $f_{3,i-1} = f_{3,i-1,a} + g_n f_{3,i-1,g}$ and $f_{4,i-1} = f_{4,i-1,a} + g_n f_{4,i-1,g}$. 

As a consequence, condition \eqref{eq:semi-discrete_local_energy_conservation/conservation_condition/2} can be split into the following four separate conditions by collecting terms featuring $g_{n}$ and those not featuring $g_{n}$, and by using the functional dependencies assumed for the fluxes: 
\begin{subequations}
\begin{align}
\begin{multlined}
-  \left\llbracket \frac{1}{2 }  \frac{q_{3,i}^2}{  \overline{q}_{1,i}^2 }  \right\rrbracket \overline{f}_{1,i}  + \left\llbracket  \frac{q_{3,i}}{\overline{q}_{1,i}}  \right\rrbracket f_{3,i,a}
\end{multlined} 
&= 0, \label{eq:semi-discrete_local_energy_conservation/specific_energy_1/split_condition_upper_no_g/neat_notation} \\
\begin{multlined}
-  \left\llbracket \frac{1}{2}  \frac{q_{4,i}^2}{  \overline{q}_{2,i}^2 }   \right\rrbracket \overline{f}_{2,i} + \left\llbracket \frac{q_{4,i}}{\overline{q}_{2,i}}   \right\rrbracket f_{4,i,a}
\end{multlined}
&= 0, \label{eq:semi-discrete_local_energy_conservation/specific_energy_1/split_condition_lower_no_g/neat_notation} \\
\begin{multlined}
\overline{ \left( \left\llbracket  g_n \left( H- H_{U,i} \right)  \right\rrbracket f_{1,i} \right)}
+ \left\llbracket  \frac{q_{3,i}}{\overline{q}_{1,i}} \right\rrbracket g_n f_{3,i,g}
\end{multlined} 
&= 
\begin{multlined}
-  \left\llbracket \rho_U g_n  \overline{\widehat{H}}_{U,i}  \frac{q_{3,i}}{\overline{q}_{1,i}}  \right\rrbracket,
\end{multlined} \label{eq:semi-discrete_local_energy_conservation/specific_energy_1/split_condition_upper_with_g/neat_notation} \\
\begin{multlined}
\overline{ \left( \left\llbracket  g_n H_{L,i}  \right\rrbracket f_{2,i} \right)}
 +  \left\llbracket  \frac{q_{4,i}}{\overline{q}_{2,i}}   \right\rrbracket g_n f_{4,i,g}
\end{multlined}  
&= 
\begin{multlined}
 - \left\llbracket \rho_L g_n  \overline{\widehat{H}}_{L,i} \frac{q_{4,i}}{\overline{q}_{2,i}} \right\rrbracket.
\end{multlined}  \label{eq:semi-discrete_local_energy_conservation/specific_energy_1/split_condition_lower_with_g/neat_notation}
\end{align}
 \label{eq:semi-discrete_local_energy_conservation/specific_energy_1/split_condition_lower_with_g_all_4/neat_notation}
\end{subequations}
These conditions have been obtained analogously to their continuous equivalents \eqref{eq:continuous_energy_conservation/specific_energy_1/split_condition_all_4}. In the continuous case, the fluxes were known and these conditions were satisfied by construction. In the discrete case, these conditions will be used in the next section to determine the numerical fluxes.  

\subsection{Derivation of energy-conserving numerical fluxes for the TFM}
\label{ssec:semi-discrete_local_energy_conservation/derivation_of_numerical_fluxes}

System \eqref{eq:semi-discrete_local_energy_conservation/specific_energy_1/split_condition_lower_with_g_all_4/neat_notation} is a system of four equations for six unknowns. To find a solution we assume, based on the continuous expression, that
\begin{equation}
\boxed{ f_{1,i-1/2} = \frac{q_{3,i-1/2}}{\Delta s} \quad \text{and} \quad f_{2,i-1/2} = \frac{q_{4,i-1/2}}{\Delta s}. }
 \label{eq:semi-discrete_local_energy_conservation/specific_energy_1/f_1_and_f_2}
\end{equation}
This choice is motivated by the fact that it requires no interpolation, and moreover is such that the discrete Poisson equation \eqref{eq:discrete_pressure_Poisson} follows naturally from the discrete volumetric flow constraint \eqref{eq:discrete_volumetric_flow_constraint} (which is used in our time integration method \cite{SanderseVeldman2019}).


Substituting $f_{1,i-1/2}$ in  \eqref{eq:semi-discrete_local_energy_conservation/specific_energy_1/split_condition_upper_no_g/neat_notation} and  $f_{2,i-1/2}$ in \eqref{eq:semi-discrete_local_energy_conservation/specific_energy_1/split_condition_lower_no_g/neat_notation} yields directly
\begin{equation}
 \boxed{ f_{3,i,a} 
 = \frac{1}{\Delta s}   \overline{ \left( \frac{q_{3,i}}{  \overline{q}_{1,i} } \right) }  \overline{q}_{3,i} \quad \text{and} \quad f_{4,i,a} 
 = \frac{1}{\Delta s}   \overline{ \left( \frac{q_{4,i}}{  \overline{q}_{2,i} } \right) }  \overline{q}_{4,i}. }
  \label{eq:semi-discrete_local_energy_conservation/specific_energy_1/f_3_i_a}
\end{equation}
To get the gravity component of $f_{3,i}$ and $f_{4,i}$, substitution of $f_{2,i-1/2}$ in \eqref{eq:semi-discrete_local_energy_conservation/specific_energy_1/split_condition_lower_with_g/neat_notation} leads to
\begin{equation*}
  \frac{1}{2} \left\llbracket H_{L,i-1/2} \right\rrbracket  \frac{q_{4,i-1/2}}{\Delta s}  + \frac{1}{2} \left\llbracket  H_{L,i+1/2} \right\rrbracket  \frac{q_{4,i+1/2}}{\Delta s} +  \left\llbracket  \frac{q_{4,i}}{\overline{q}_{2,i}}   \right\rrbracket f_{4,i,g} 
 =     \rho_L   \overline{\widehat{H}}_{L,i-1/2} \frac{q_{4,i-1/2}}{\overline{q}_{2,i-1/2}} - \rho_L   \overline{\widehat{H}}_{L,i+1/2} \frac{q_{4,i+1/2}}{\overline{q}_{2,i+1/2}}. 
 \end{equation*}
After significant rewriting, this yields the following expression for the gravity component of $f_{4,i}$:
\begin{equation}
f_{4,i,g} 
 = - \rho_L \widehat{H}_{L,i} -    \overline{\left[ \left( \rho_L \frac{\left\llbracket  \widehat{H}_{L,i} \right\rrbracket}{\overline{q}_{2,i}} +  \frac{\left\llbracket  H_{L,i} \right\rrbracket}{\Delta s} \right) q_{4,i} \right]} \left\llbracket  \frac{q_{4,i}}{\overline{q}_{2,i}} \right\rrbracket^{-1}, 
\label{eq:semi-discrete_local_energy_conservation/specific_energy_1/general_result_f_4_i_g}
\end{equation}
and a similar expression holds for $f_{3,i,g}$.

The first term on the left-hand side is easily recognized as the discrete counterpart of $-\rho_{L}  \widehat{H}_{L}$. 
In order for the discrete expression to be practical and match the continuous expression, the second term must vanish, and we require the following conditions to be satisfied:
\begin{equation}
\left\llbracket  \widehat{H}_{U,i+1/2} \right\rrbracket =   \frac{\overline{q}_{1,i+1/2}}{\rho_U \Delta s} \left\llbracket H_{U,i+1/2} \right\rrbracket,
\qquad 
\left\llbracket  \widehat{H}_{L,i-1/2} \right\rrbracket = -  \frac{\overline{q}_{2,i-1/2}}{\rho_L \Delta s} \left\llbracket H_{L,i-1/2} \right\rrbracket.
\label{eq:semi-discrete_local_energy_conservation/specific_energy_1/check_of_better_deduction/geometric_condition/together} 
\end{equation}
In the continuous case a continuous version of these conditions, given by \eqref{eq:continuous_energy_conservation/entropy_potential/geometric_condition_q_together_s}, is also required, and these can be shown to be satisfied exactly via manipulation of the continuous derivatives.  
The same manipulation is not possible with discrete jumps, so that in the discrete case these conditions are not satisfied in general, and the second term in  \eqref{eq:semi-discrete_local_energy_conservation/specific_energy_1/general_result_f_4_i_g} does not generally vanish.
This means that we cannot obtain a practical energy-conserving discretization for arbitrary geometries (at least not with the conventional staggered-grid finite volume method that we have employed).

Even though conditions \eqref{eq:semi-discrete_local_energy_conservation/specific_energy_1/check_of_better_deduction/geometric_condition/together} are not generally exactly satisfied in the discrete case, we can show that they are approximately satisfied for arbitrary duct geometries, and that they are exactly satisfied for  specific geometries such as a channel.
This can be shown by evaluating both sides of  \eqref{eq:semi-discrete_local_energy_conservation/specific_energy_1/check_of_better_deduction/geometric_condition/together} using Taylor series.
We expand $\widehat{H}_{L,i-1}$ and $H_{L,i-1}$ into Taylor series around $A_L=A_{L,i}$, and expand $A_{L,i-1}$ around $s=s_{i}$.
These Taylor series are combined to obtain expressions for $\left\llbracket \widehat{H}_{L,i-1/2} \right\rrbracket$,  $\left\llbracket H_{L,i-1/2} \right\rrbracket$, and $\left\llbracket A_{L,i-1/2} \right\rrbracket$.
With these expressions the left-hand side of \eqref{eq:semi-discrete_local_energy_conservation/specific_energy_1/check_of_better_deduction/geometric_condition/together} evaluates to
\begin{multline}
  \left\llbracket \widehat{H}_{L,i-1/2} \right\rrbracket = 
-\left(\d{\widehat{H}_{L}}{A_{L}}\right)_{i} \left( A_{L,i-1} - A_{L,i} \right) - \frac{1}{2}  \left(\dd{\widehat{H}_{L}}{A_{L}}\right)_{i} \left( A_{L,i-1} - A_{L,i} \right)^2 \\
- \frac{1}{6}  \left(\ddd{\widehat{H}_{L}}{A_{L}}\right)_{i} \left( A_{L,i-1} - A_{L,i} \right)^3 
 + O(\Delta s^4), 
  \label{eq:semi-discrete_local_energy_conservation/specific_energy_1/better_deduction_condition_check/LHS_1} 
\end{multline}
where $(.)_{i}$ indicates $(.)$ evaluated at $A_{L,i}$. The right-hand side of \eqref{eq:semi-discrete_local_energy_conservation/specific_energy_1/check_of_better_deduction/geometric_condition/together} evaluates to
 \begin{multline}
- \frac{\overline{q}_{2,i-1/2}}{\rho_L \Delta s} \left\llbracket H_{L,i-1/2} \right\rrbracket
=
\frac{1}{2} \left(\d{H_{L}}{A_{L}}\right)_{i} \left( A_{L,i-1}^2 - A_{L,i}^2 \right)  
+ \frac{1}{4} \left(\dd{H_{L}}{A_{L}}\right)_{i} \left( A_{L,i-1} + A_{L,i} \right) \left( A_{L,i-1} - A_{L,i} \right)^2 \\
+ \frac{1}{12} \left(\ddd{H_{L}}{A_{L}}\right)_{i} \left( A_{L,i-1} + A_{L,i} \right) \left( A_{L,i-1} - A_{L,i} \right)^3 + O(\Delta s^4). 
 \label{eq:semi-discrete_local_energy_conservation/specific_energy_1/better_deduction_condition_check/RHS_1} 
\end{multline}
At this point we apply relation \eqref{eq:appendix/derivative_of_H_hat_relation_to_derivative_of_H} from \autoref{sec:geometric_details} to the discrete quantities used here:
\begin{equation*}
\left(\d{\widehat{H}_{L}}{A_{L}}\right)_{i} = - A_{L,i} \left(\d{H_{L}}{A_{L}}\right)_{i},
\end{equation*}
and from this we can derive
\begin{equation*}
\left(\dd{\widehat{H}_{L}}{A_{L}}\right)_{i} =   - \left(\d{H_{L}}{A_{L}}\right)_{i} - A_{L,i} \left(\dd{H_{L}}{A_{L}}\right)_{i}, 
\quad \text{and} \quad 
\left(\ddd{\widehat{H}_{L}}{A_{L}}\right)_{i} =   - 2 \left(\dd{H_{L}}{A_{L}}\right)_{i} - A_{L,i} \left(\ddd{H_{L}}{A_{L}}\right)_{i}.
\end{equation*}
Substitution of these relations in \eqref{eq:semi-discrete_local_energy_conservation/specific_energy_1/better_deduction_condition_check/LHS_1}, and comparison of the result to \eqref{eq:semi-discrete_local_energy_conservation/specific_energy_1/better_deduction_condition_check/RHS_1} yields
\begin{equation}
\begin{split}
 \left\llbracket \widehat{H}_{L,i-1/2} \right\rrbracket &=
 - \frac{\overline{q}_{2,i-1/2}}{\rho_L \Delta s} \left\llbracket H_{L,i-1/2} \right\rrbracket + \frac{1}{12} \left(\dd{H_{L}}{A_{L}}\right)_{i}  \left( A_{L,i-1} - A_{L,i} \right)^3  + O(\Delta s^4).
 \end{split}
 \label{eq:semi-discrete_local_energy_conservation/geometry_error}
\end{equation}
This derivation can be carried out with similar results for the upper fluid. 

These relations show that for arbitrary duct geometries, the geometric conditions \eqref{eq:semi-discrete_local_energy_conservation/specific_energy_1/check_of_better_deduction/geometric_condition/together} are satisfied only approximately in the discrete case.
This stands in contrast to the continuous case, where the equivalent geometric conditions are satisfied exactly (for arbitrary geometries). 

Fortunately, for a 2D channel geometry $\mathrm{d} H_{L}/\mathrm{d} A_{L} = 1$ and $\mathrm{d}^2 H_{L}/\mathrm{d} A_{L}^2 = 0$, and all higher order derivatives are zero, so in this case \eqref{eq:semi-discrete_local_energy_conservation/specific_energy_1/check_of_better_deduction/geometric_condition/together} is exactly satisfied.
This means that the 2D channel geometry is an important special case for which we obtain the following numerical fluxes:
\begin{equation}
\boxed{
f_{3,i,g} =  - \rho_U \widehat{H}_{U,i}  
\quad \text{and} \quad
f_{4,i,g} =  - \rho_L \widehat{H}_{L,i}.} \label{eq:semi-discrete_local_energy_conservation/specific_energy_1/channel_result_f_4_i_g} 
\end{equation}
These fluxes are energy-conserving for other geometries with $\mathrm{d}^2 H_{L}/\mathrm{d} A_{L}^2 = 0$, but not for geometries with curved sides, such as the pipe geometry.

The final collection of energy-conserving numerical fluxes is given by \eqref{eq:semi-discrete_local_energy_conservation/specific_energy_1/f_1_and_f_2}, \eqref{eq:semi-discrete_local_energy_conservation/specific_energy_1/f_3_i_a}, and \eqref{eq:semi-discrete_local_energy_conservation/specific_energy_1/channel_result_f_4_i_g}.
Of these, \eqref{eq:semi-discrete_local_energy_conservation/specific_energy_1/f_1_and_f_2} and \eqref{eq:semi-discrete_local_energy_conservation/specific_energy_1/channel_result_f_4_i_g} are locally exact, and \eqref{eq:semi-discrete_local_energy_conservation/specific_energy_1/f_3_i_a} involves second order accurate central interpolation.
Together they form the numerical flux vector
\begin{equation}
\v{f}_{i-1/2}(\v{q}_{i-2},\v{q}_{i-1},\v{q}_i) 
=
\begin{bmatrix}
\frac{q_{3,i-1/2}}{\Delta s} \\
\frac{q_{4,i-1/2}}{\Delta s} \\
\frac{1}{\Delta s}   \overline{ \left( \frac{q_{3,i-1}}{  \overline{q}_{1,i-1} } \right) }  \overline{q}_{3,i-1}  - \rho_U g_n \widehat{H}_{U,i-1} \\
\frac{1}{\Delta s}   \overline{ \left( \frac{q_{4,i-1}}{  \overline{q}_{2,i-1} } \right) }  \overline{q}_{4,i-1} - \rho_L g_n \widehat{H}_{L,i-1}
\end{bmatrix}
=\begin{bmatrix}
\rho_U \overline{A}_{U,i-1/2} u_{U,i-1/2} \\
\rho_L \overline{A}_{L,i-1/2} u_{L,i-1/2}  \\
\rho_U \overline{u}_{U,i} \overline{ \left(   \overline{A}_{U,i-1} u_{U,i-1} \right)} - \rho_U g_n \widehat{H}_{U,i-1} \\
\rho_L \overline{u}_{L,i} \overline{ \left(   \overline{A}_{U,i-1} u_{U,i-1} \right)} - \rho_L g_n \widehat{H}_{L,i-1}
\end{bmatrix}. 
\label{eq:semi-discrete_local_energy_conservation/specific_energy_1/numerical_flux_vector} 
\end{equation}
Here the flux is rendered in terms of primitive variables only for ease of interpretation; the implementation of the numerical flux (and of the discrete energy) in the numerical code is completely in terms of the conservative variables.

\rmk{The difficulty to satisfy condition \eqref{eq:semi-discrete_local_energy_conservation/specific_energy_1/check_of_better_deduction/geometric_condition/together} for arbitrary cross-sectional geometries is not dependent on the choice of $\psi_{i-1/2}$, nor is it due to the interpolation of the potential energy to the velocity grid points (as needed on a staggered grid). This is shown in  \autoref{sec:global_discrete_energy_conservation} by applying a global energy analysis.}


\rmk{The proposed discrete energy \eqref{eq:semi-discrete_local_energy_conservation/specific_energy_1/energy_definition} is a consistent approximation to \eqref{eq:continuous_energy_conservation/conservative/e_definition_1} which is conserved by the numerical fluxes given by \eqref{eq:semi-discrete_local_energy_conservation/specific_energy_1/numerical_flux_vector}.
However, it is not unique. For example, an alternative definition is
\begin{equation}
e_{i}(\v{q}_{i},\v{q}_{i+1})   =  \rho_U g_n \widetilde{H}_{U,i} \Delta s+  \rho_L g_n \widetilde{H}_{L,i} \Delta s + \frac{1}{2} \overline{ \left(\frac{q_{3,i}^2}{ \overline{q}_{1,i}  } \right)} + \frac{1}{2 } \overline{\left(\frac{q_{4,i}^2}{ \overline{q}_{2,i}  } \right)}.
\label{eq:semi-discrete_local_energy_conservation/pressure_grid/energy_definition}
\end{equation}
In this formulation the energy is defined on the pressure grid, and the energy conservation conditions and local energy conservation equation can be adapted to accommodate for this.
With a similar change in the entropy potential, it is again possible to derive a set of energy-conserving numerical fluxes, which turn out to be the same as those given by \eqref{eq:semi-discrete_local_energy_conservation/specific_energy_1/numerical_flux_vector}.
As the issue of the geometric relations also persists with this choice, there seems no clear advantage over our proposed formulation.}


\section{Numerical experiments} \label{sec:simulations}

We perform numerical experiments for a 2D channel geometry, with the goal of verifying conservation of the discrete global energy, as discussed in \autoref{sec:semi-discrete_local_energy_conservation/outline}:
\begin{equation*}
\d{E_{h}}{t} = 0.
\end{equation*}
The model for which we perform the experiments will not include source terms such as wall friction and interface friction, or diffusion, since these would lead to dissipation of energy in the continuous analysis.
The test cases are chosen such that no discontinuities appear, for which the continuous analysis is invalid, since this would also necessitate dissipation of energy. 
Furthermore, the numerical experiments performed in this section will all be in the `well-posed regime' of the TFM, meaning that the initial conditions are chosen such that the eigenvalues of the model are real, and remain so.

We use the discretization as outlined in \autoref{sec:discretization}, with the numerical fluxes given by \eqref{eq:semi-discrete_local_energy_conservation/specific_energy_1/numerical_flux_vector}. The vector $\v{d}_i$ of the pressure term is given by \eqref{eq:semi-discrete_local_energy_conservation/specific_energy_1/pressure_d_discretization}. 
We noted earlier that the scheme is spatially exactly energy-conserving, but not temporally. However, we can still obtain energy conservation by taking the time step sufficiently small. The difference between the initial energy $E_{h}^0$ and the final energy $E_{h}^{N_t}$ after $N_t$ time steps should then be in the order of the machine precision, and we shall term this difference the `energy error'.

 \subsection{Gaussian perturbation in a periodic domain}
 \label{ssec:results/gaussian}

We consider a test case with periodic boundaries, so that effectively we do not need to take the boundaries into account.
We introduce a perturbation in the hold-up $\alpha_L = A_L/A$ of the form
\begin{equation*}
\alpha_L(s) = \alpha_{L,0} + \Delta \alpha_L(s), \qquad \Delta \alpha_L(s) =  \Delta \widehat{\alpha}_L \exp \left[- \frac{1}{2}  \left( \frac{s - L/2}{\sigma} \right)^2 \right],
\label{eq:results/gaussian_perturbation_general_form}
\end{equation*}
with $ \Delta \widehat{\alpha}_L = 0.2$ and $\sigma = L/10$, and $L$ the length of the domain. 
This produces a Gaussian perturbation centered at the middle of the domain. 
The initial velocities are left at zero, which ensures exact initial satisfaction of the volumetric flow constraint \eqref{eq:governing_equations/volumetric_flux_constraint} (in fact, $Q=0$).

We use parameters similar to those used in the Thorpe experiment \cite{Thorpe1969}, as described by \cite{LopezdeBertodanoFullmerClausse2016}.
They are given by \autoref{tab:gaussian_test_case_parameters}.
The choice for a large upper fluid density is deliberate: it ensures that all terms in the expression for $e$, \eqref{eq:semi-discrete_local_energy_conservation/specific_energy_1/energy_definition}, are significant. 

\begin{table}[htb]
\small
\centering
\caption{Parameters for the Gaussian perturbation test case.}
\begin{tabular}{llll}
\toprule
Parameter & Symbol & Value & Units\\
\midrule
Lower fluid density & $\rho_L$ & $1000$ & $  \mathrm{kg} \, \mathrm{m}^{-3}$ \\
Upper fluid density & $\rho_U$ & $780$ & $ \mathrm{kg} \, \mathrm{m}^{-3}$ \\
Acceleration of gravity & $g$ & $9.8$ & $\mathrm{m} \, \mathrm{s}^{-2}$ \\
Channel inclination & $\phi$ & $0$ & degrees  \\
Domain length & $L$ & $1.83$ & $\mathrm{m}$   \\
Channel height & $H$ & $0.03$ & $ \mathrm{m}$ \\
Initial lower fluid hold-up & $\alpha_{L,0}$ & $0.5$ & $-$ \\
Initial lower fluid velocity & $u_{L,0}$ & $0$ & $\mathrm{m} \, \mathrm{s}^{-1}$ \\
Initial upper fluid velocity & $u_{U,0}$ & $0$ & $\mathrm{m} \, \mathrm{s}^{-1}$ \\
\bottomrule
\end{tabular}
\label{tab:gaussian_test_case_parameters}
\end{table}

\begin{figure}[htbp] 
\centering
\includegraphics[width=0.45\linewidth]{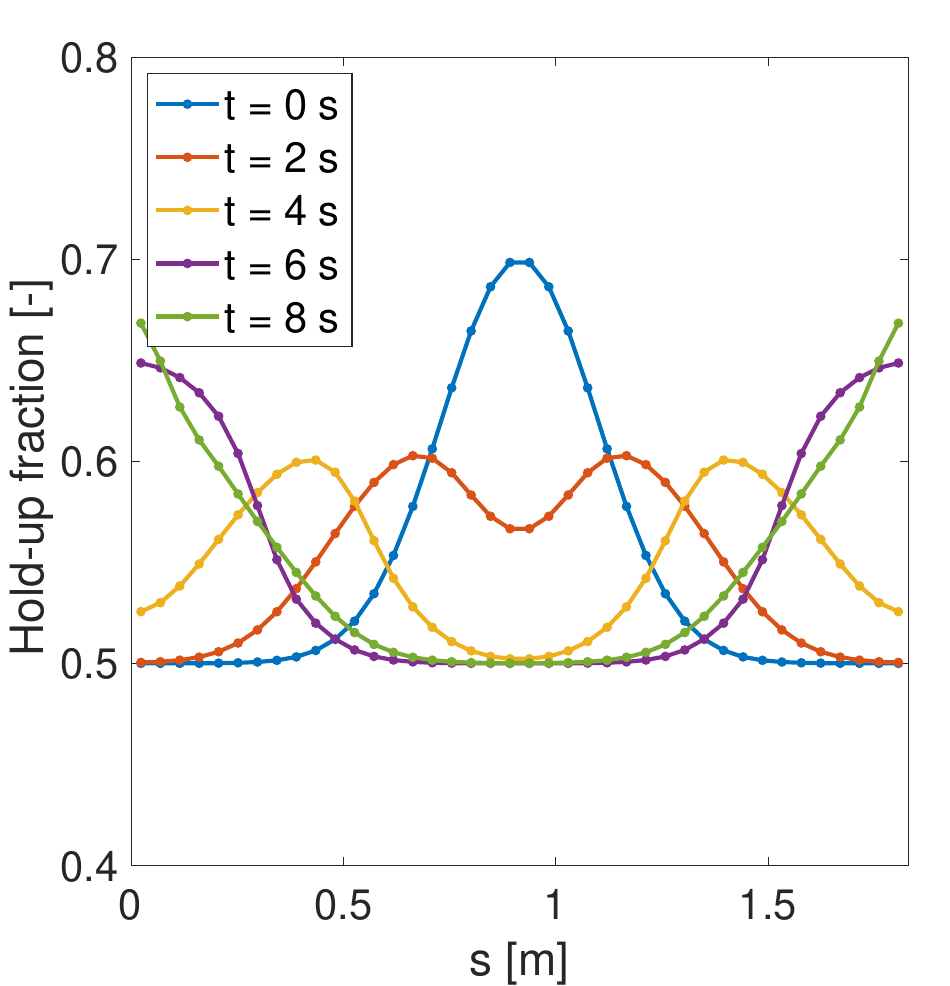} 
\includegraphics[width=0.45\linewidth]{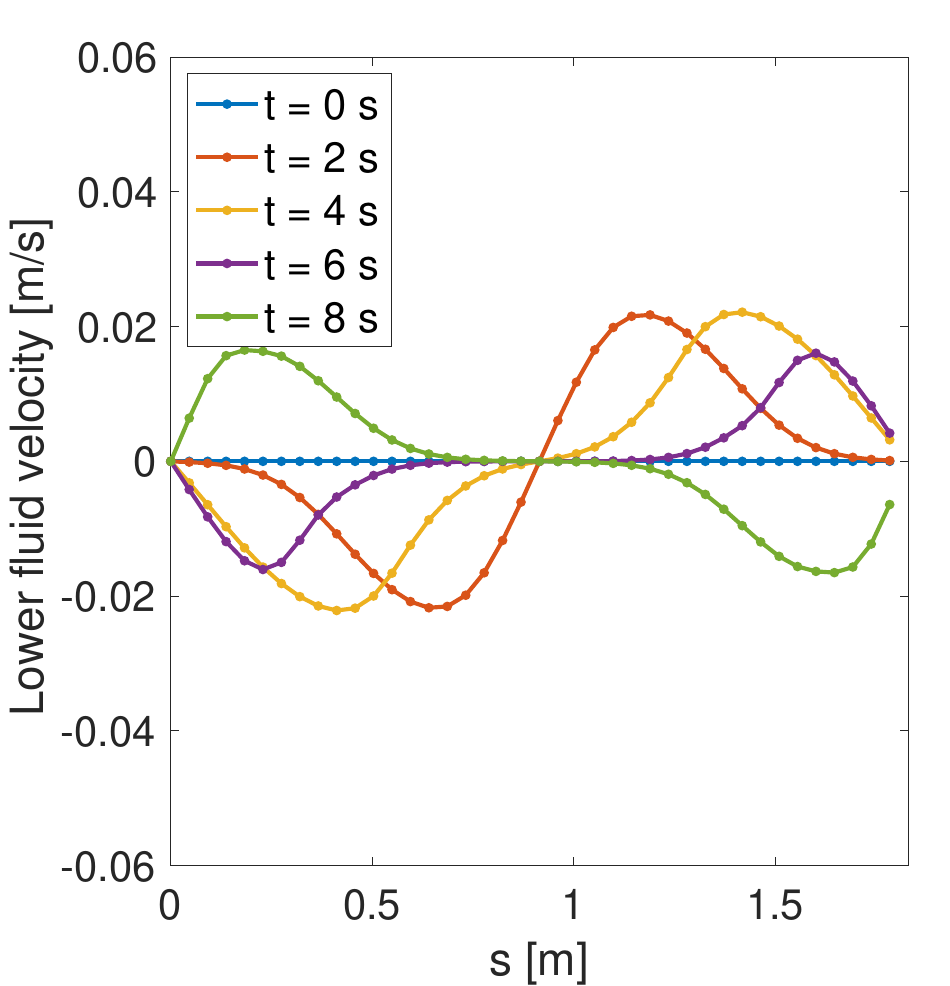} 
\caption{The initial evolution of the Gaussian perturbation, up to the point that the boundaries are met. Left: lower fluid hold-up. Right: lower fluid velocity.}
\label{fig:results/gaussian/evolution}
\end{figure}

We employ $N_{p}=N_{u}=40$ finite volumes with $\Delta s = L/N_p$ and let the simulations run until $t=30 \, \mathrm{s}$, with $\Delta t = 0.001 \, \mathrm{s}$. The perturbation splits symmetrically into a left-traveling and a right-traveling wave, which travel through the periodic boundaries, to eventually come together in the middle and reform the initial perturbation approximately.
We show the evolution of the hold-up and velocity in \autoref{fig:results/gaussian/evolution}, roughly up to the point that the waves meet at the boundaries of the domain.

\begin{figure}[hbtp] 
\centering
\includegraphics[width=0.45\linewidth]{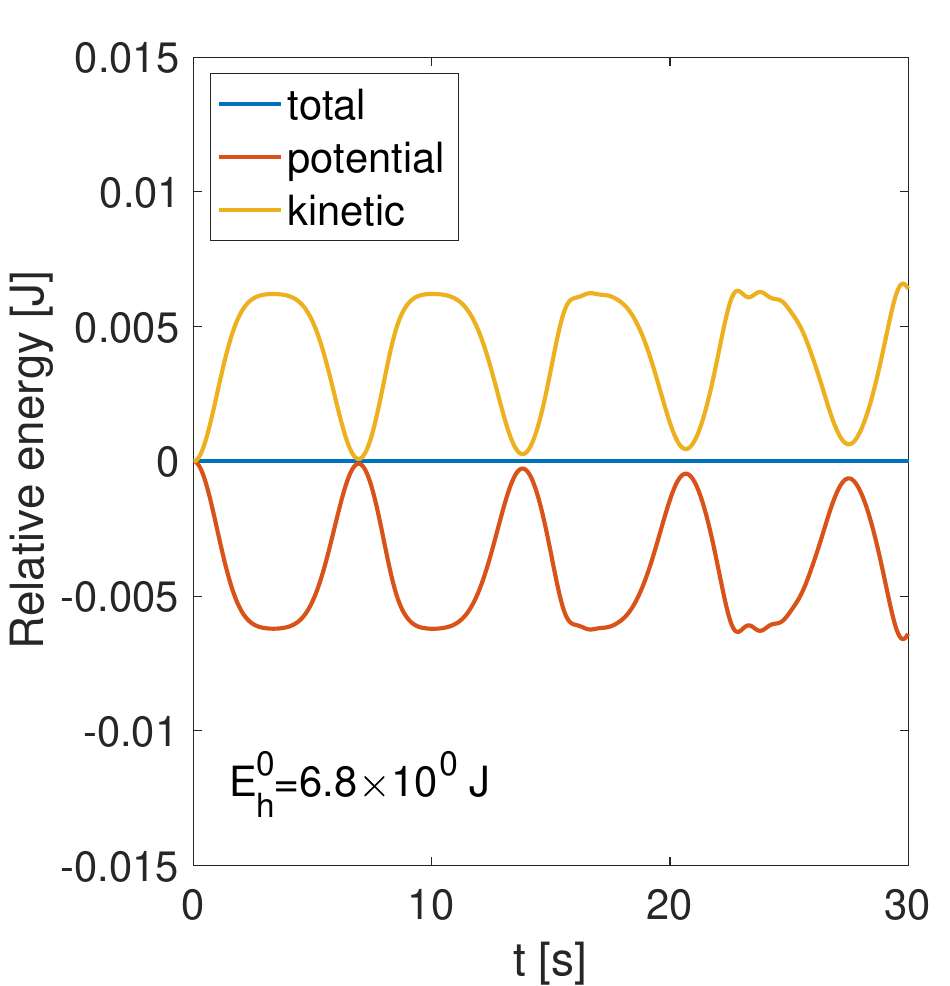} 
\includegraphics[width=0.45\linewidth]{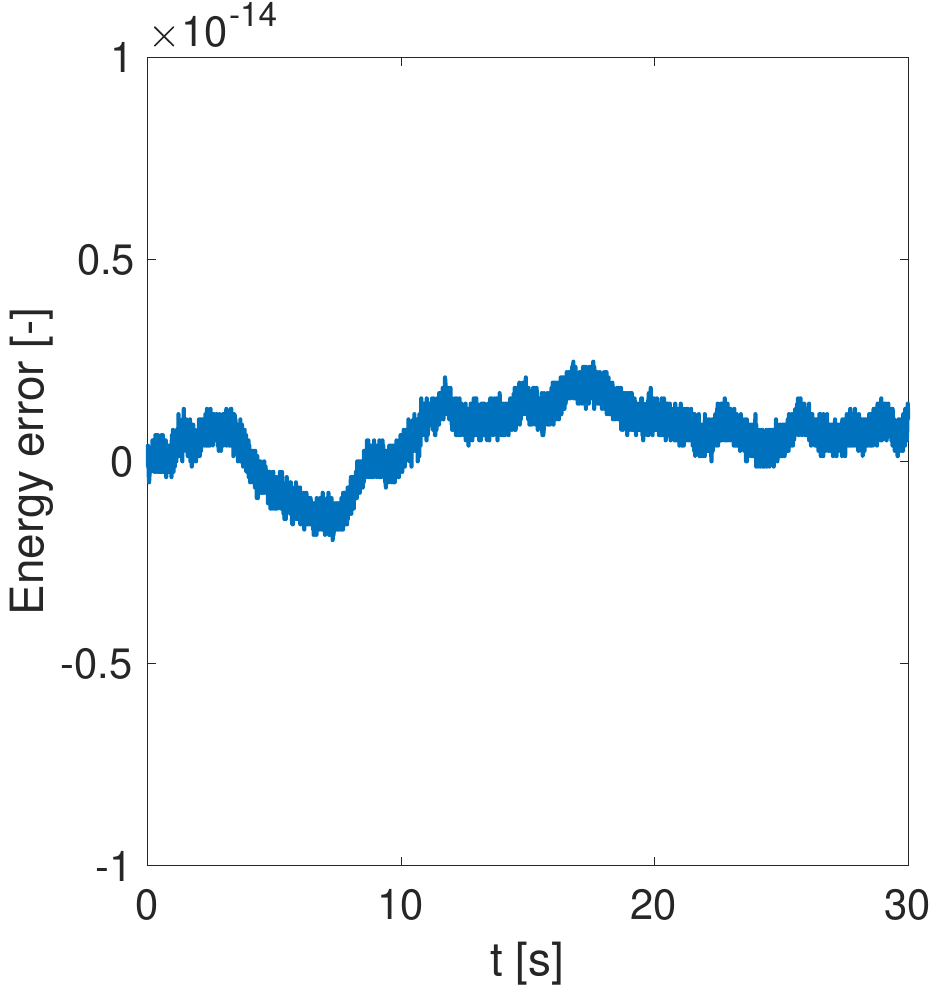}
\caption{Conserved quantities for the Gaussian perturbation test case. Left: potential, kinetic and total energy relative to their initial values. Right: $\left(E_{h}-E_h^0\right)/E_h^0$. }
\label{fig:results/gaussian/conservation_in_time_plot_N_0040}
\end{figure}

In this test case we have a significant exchange between kinetic and potential energy, which can be seen in \autoref{fig:results/gaussian/conservation_in_time_plot_N_0040} (left panel). The total energy
is conserved up to machine precision, as can be seen in the right panel of the figure. 
The mass of each phase and total momentum are also conserved, and the volume constraint and volumetric flow constraint are satisfied, up to machine precision (see also \cite{SanderseVeldman2019}). 
As time progresses, nonlinear effects start to play a role, leading to more irregular behaviour of the potential and kinetic energy as a function of time. The sum of the two stays exactly constant, confirming our theoretical derivations, and showing that our newly proposed numerical fluxes for the TFM lead indeed to an energy-conserving discretization method. 

\begin{figure}[htbp] 
\centering
\includegraphics[width=0.45\linewidth]{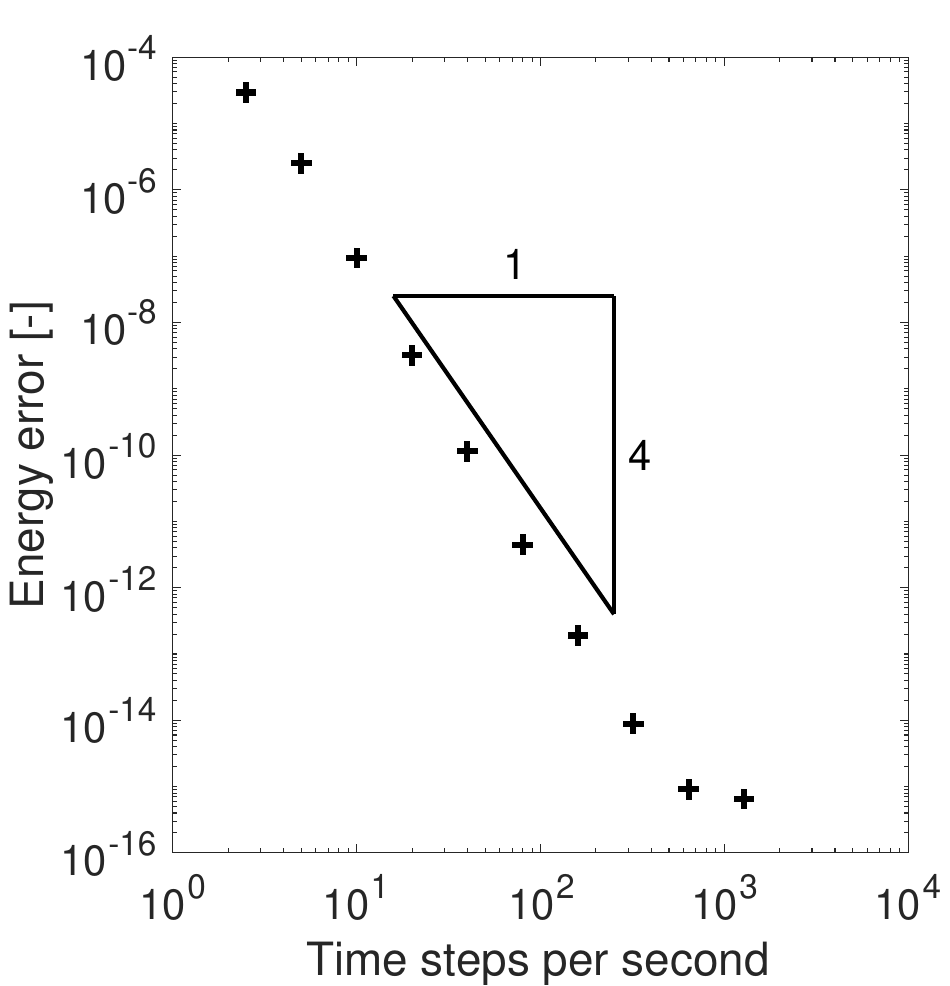} 
\caption{Convergence of the energy error $\left(E_h^{N_t}-E_h^0\right)/E_h^0$ with time step, for the Gaussian perturbation test case. 
}
\label{fig:results/gaussian/E^N-E^0}
\end{figure}

We give further evidence that the energy is conserved exactly by the spatial discretization, and limited only by a temporal error, by plotting the convergence of the energy error with refinement of the time step. 
\autoref{fig:results/gaussian/E^N-E^0} shows a fourth order convergence rate with $\Delta t$, in agreement with the fourth order accuracy of the Runge-Kutta time integration method. 
The convergence continues up to machine precision, which is reached around $\Delta t=0.001 \, \mathrm{s}$, as was used for the results in \autoref{fig:results/gaussian/conservation_in_time_plot_N_0040}, confirming that the spatial discretization conserves energy up to machine precision.

\FloatBarrier

\subsection{Sloshing in a closed tank}
\label{ssec:results/sloshing}

\begin{figure}[htbp] 
\centering
\includegraphics[width=0.45\linewidth]{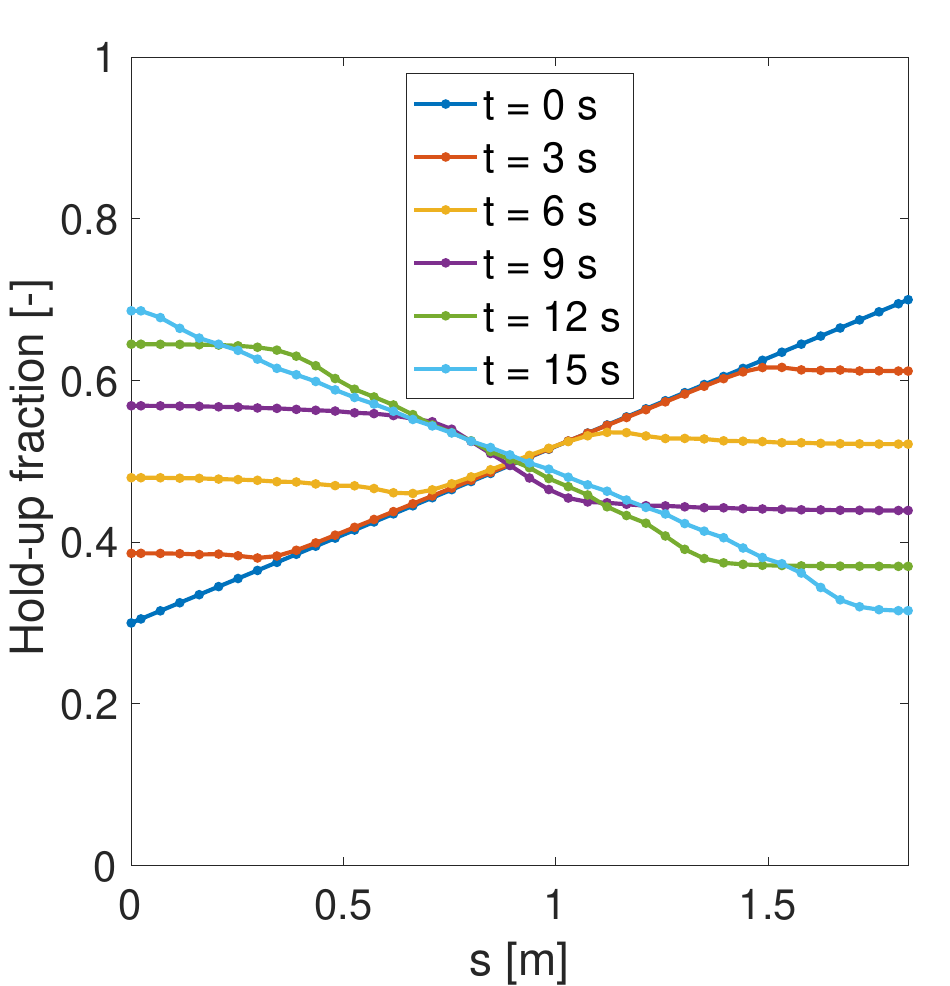} 
\includegraphics[width=0.45\linewidth]{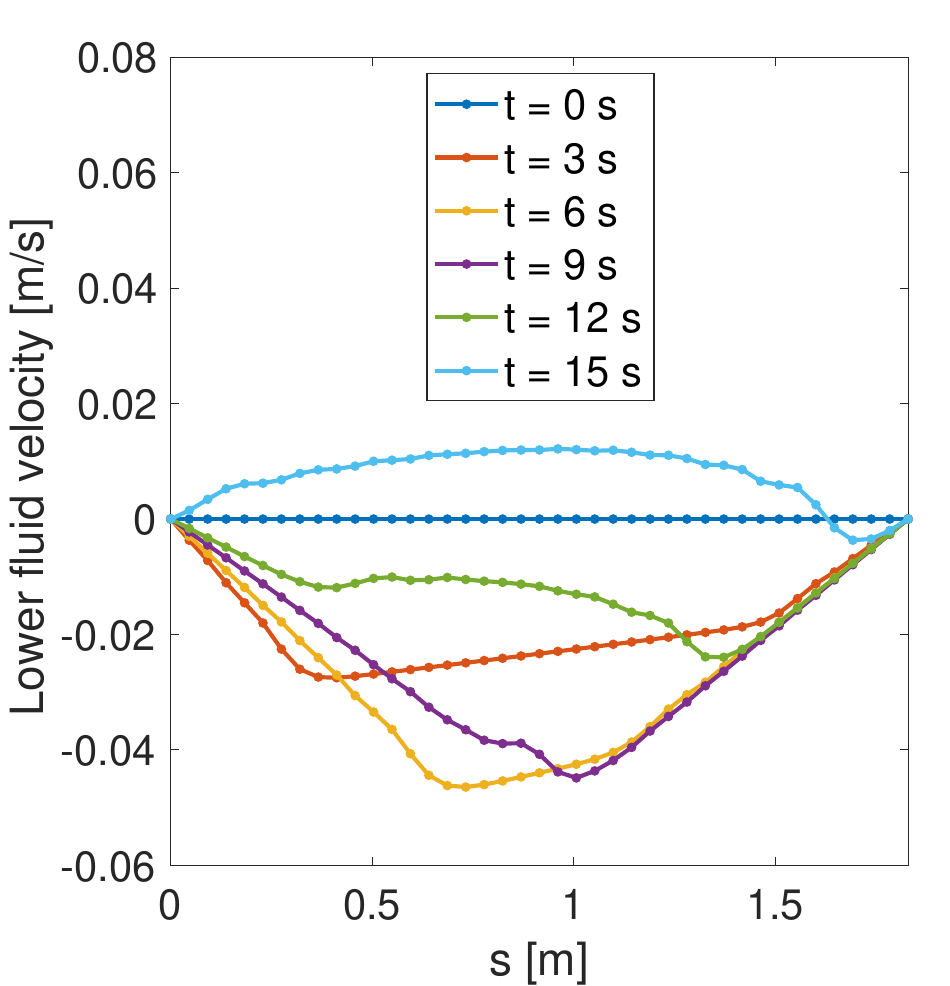} 
\caption{The initial evolution of the sloshing simulation, approximately up to the point that the lower fluid reaches its maximum height at the left boundary. Left: lower fluid hold-up. Right: lower fluid velocity.}
\label{fig:results/sloshing/evolution}
\end{figure}

We now consider a test case with closed (solid-wall) boundaries, for which energy conservation is expected to hold because the fluxes $h$ and $j$ involve multiplication with $q_{3}$ and $q_{4}$, which are zero at the boundaries. The test case features a closed rectangular tank in which the two fluids are brought out of equilibrium, so that sloshing occurs.  The parameters are identical to those of the previous test case, see \autoref{tab:gaussian_test_case_parameters}, except that the initial condition for the hold-up perturbation is different. It is given by
\begin{equation*}
\alpha_L(s) = \alpha_{L,0} + \Delta \alpha_L(s), \qquad \Delta \alpha_L(s) =  \Delta \widehat{\alpha}_L  \frac{s-L/2}{L/2},
\label{eq:sloshing_perturbation_general_form}
\end{equation*}
with $\Delta \widehat{\alpha}_L = 0.2$.
This yields a straight slanted interface, with $\alpha_L=0.3$ at the left boundary and $\alpha_L=0.7$ at the right boundary: see \autoref{fig:results/sloshing/evolution}. 

This is not a typical sloshing case, since the TFM was designed to model long-wavelength phenomena, 
and indeed we have taken $L \gg H$.
Therefore we are not able to explicitly capture typical sloshing phenomena such as wave breaking.
However, the effect of such small-scale phenomena on the averaged flow may be included in the model via closure terms \cite{Holmas2010}. 
With accurate closure terms, the TFM can closely match DNS results, as shown in \cite{BuistSandersevanHalderEtAl2019}. 
This is not included here, as this would lead to dissipation of energy and not allow us to show the energy-conserving properties of our proposed numerical discretization.

Like in the first test case, initially the total energy of the system consists of only potential energy. Under the presence of gravity (via the level gradient terms) the interface starts to flatten, which is achieved via a right-running and a left-running wave, that emanate from the left and right boundary, respectively. Around $t=7 \, \mathrm{s}$ the interface is almost completely flat, and all potential energy has been converted into kinetic energy, and the interface starts to slant (`slosh') again in the opposite direction. 
\autoref{fig:results/sloshing/evolution} shows this behavior up to approximately the point that the lower fluid reaches its maximum height at the left boundary.
Note that the evolution of the hold-up fraction is not exactly symmetric, amongst others because the wave speed in the `deep' part is different from the wave speed in the `shallow' part.
Also in this test case, the mass of each phase is conserved up to machine precision, but there is a (physical) inflow of momentum at the boundaries, due to the level gradient terms.

\begin{figure}[htbp] 
\centering
\includegraphics[width=0.45\linewidth]{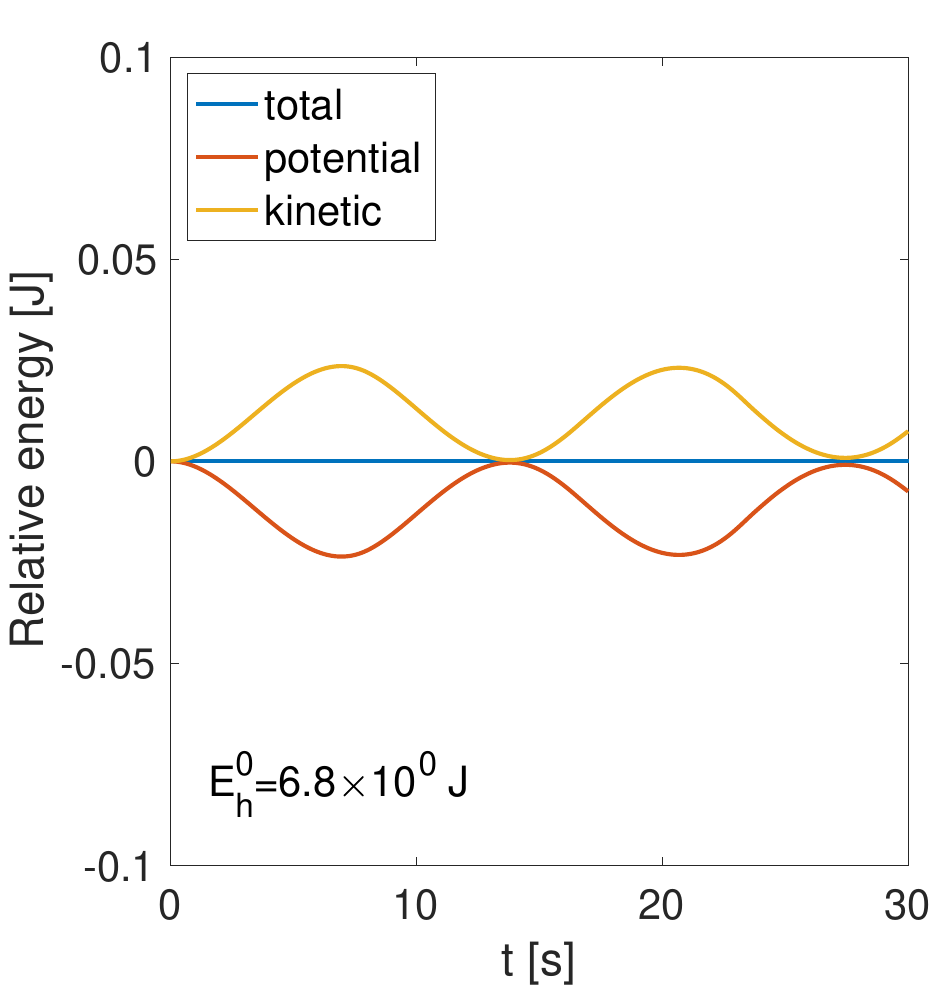} 
\includegraphics[width=0.45\linewidth]{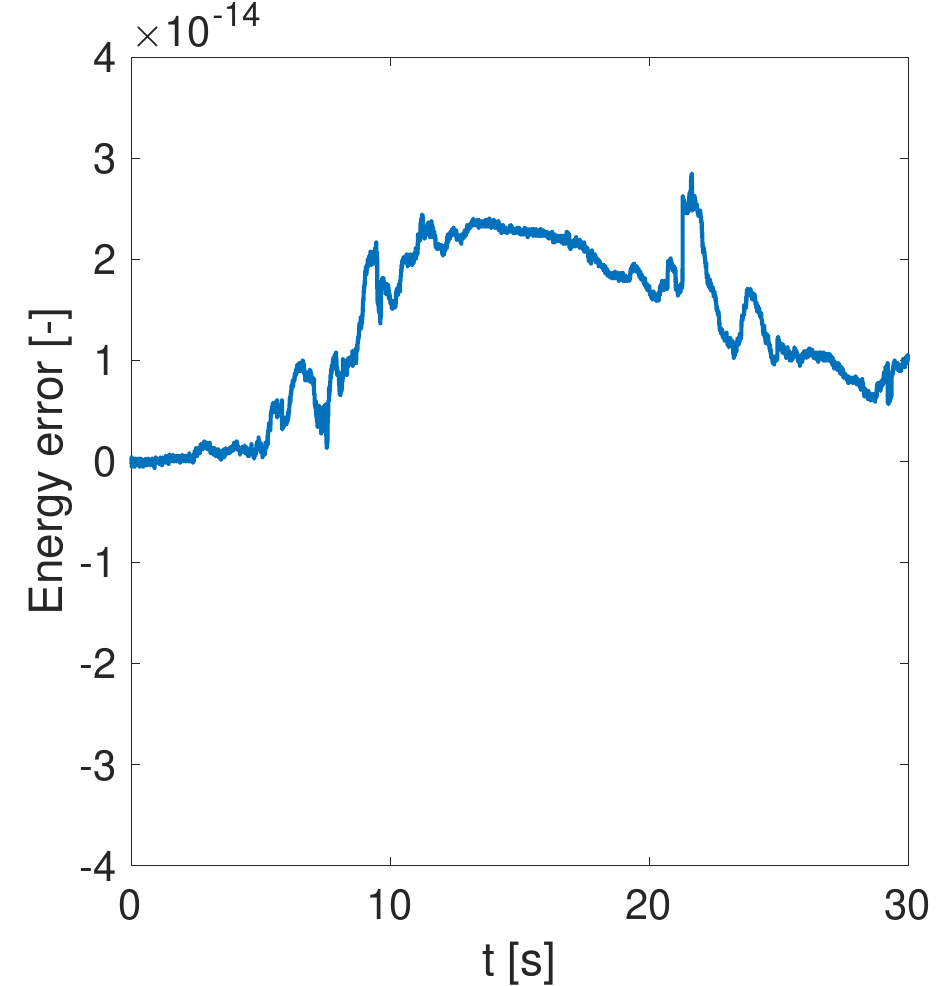}
\caption{Conserved quantities for the sloshing test case. Left: potential, kinetic and total energy relative to their initial values. Right: $\left(E_{h}-E_h^0\right)/E_h^0$. }
\label{fig:results/sloshing/conservation_in_time_plot_N_0040}
\end{figure}

\begin{figure}[htbp] 
\centering
\includegraphics[width=0.45\linewidth]{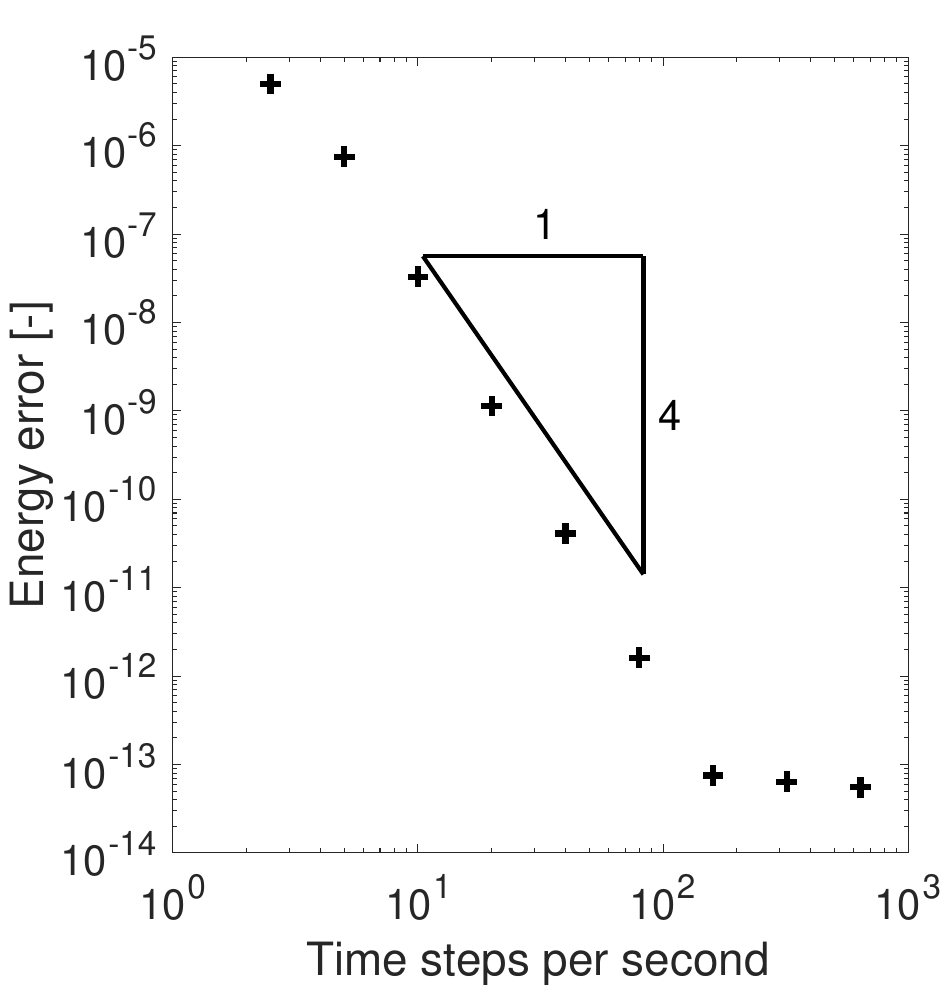} 
\caption{Convergence of the energy error $\left(E_h^{N_t}-E_h^0\right)/E_h^0$ with time step, for the sloshing test case. 
}
\label{fig:results/sloshing/E^N-E^0}
\end{figure}

\autoref{fig:results/sloshing/conservation_in_time_plot_N_0040} shows the exchange of potential and kinetic energy as a function of time. Similar to the previous test case, exact energy conservation is achieved with our proposed spatial discretization, if the time step is fine enough (here $\Delta t = 0.005 \, \mathrm{s}$, and $N_p=40$). 
If the time step is not fine enough, a (small) energy error is made, which converges with fourth order upon time step refinement, as is shown in \autoref{fig:results/sloshing/E^N-E^0}.
The ability to conserve energy in this closed system is an important step in order to obtain fidelity in the simulation results. Non-energy-conserving schemes, \eg schemes that dissipate energy, would introduce artificial (numerical) damping of the sloshing movement and incur a loss in the liquid height reached at the boundaries. In a way, the sloshing movement can be compared to a moving pendulum \cite{OckendonOckendon2017}, for which it is well-known that conservation of the total energy (the Hamiltonian) is an important property that should be mimicked upon discretization in order to achieve realistic long-time behavior.

\FloatBarrier

\subsection{Traveling wave} \label{ssec:results/kelvin-helmholtz} 

Finally, we perform a test case with a traveling wave in a periodic domain.
The flow is uni-directional and stratified, with a velocity and density difference between the two fluids.
We consider a steady base state, upon which a small periodic perturbation is introduced, of which we study the evolution in time.
This case is similar to test cases examining the Kelvin-Helmholtz instability, such as in \cite{LiaoMeiKlausner2008, SanderseVeldman2019}.
However, here the perturbation will be stable since the flow is inviscid and in the (linearly) well-posed regime.

Most of the parameters are again identical to those given by  \autoref{tab:gaussian_test_case_parameters}, but the initial conditions for the hold-up and the fluid velocities are different. 
We set $\alpha_{L,0}=0.4$ and $u_{L,0}=1$. 
For $u_U$ we take $u_{U,0}=1.187$ (which is the value that would result in a steady flow with wall and interface friction\footnote{For this we take the Churchill friction model \cite{Churchill1977} with viscosities of $\mu_U = 1.5 \cdot 10^{-3}$ $\mathrm{kg} \, \mathrm{m}^{-1} \, \mathrm{s}^{-1}$ and $\mu_L = 1 \cdot 10^{-3}$ $\mathrm{kg} \, \mathrm{m}^{-1} \, \mathrm{s}^{-1}$.}).

 
In order to construct an initial perturbation that results in a traveling wave, we conduct a linear stability analysis of the TFM \cite{LiaoMeiKlausner2008}.
The analysis is conducted in terms of its primitive variables in the form
\begin{equation}
\v{w}^T = 
\begin{bmatrix}
\alpha_L &  u_L & u_G &  p
\end{bmatrix}.
\label{eq:perturbation_amplitude_general_form}
\end{equation}
As exact solutions we obtain waves of the form
\begin{equation}
\Delta \v{w} = \mathrm{Re} \left(\Delta \widehat{\v{w}} \exp \left[i \left(\omega t - k s\right)\right]\right),
\label{eq:perturbation_general_form}
\end{equation}
with $\Delta \widehat{\v{w}}$ the amplitude of the perturbation in each variable.
The relative amplitudes in $\Delta \widehat{\v{w}}$ are such that $\Delta \widehat{\v{w}}$ is an eigenvector corresponding to one of two dispersion relations $\omega(k)$. 

The initial perturbation is defined as \eqref{eq:perturbation_general_form}, with $t=0$.
We take a wavenumber of  $k=2\pi/L \, \mathrm{m}^{-1}$ and calculate the corresponding angular frequencies, of which one is selected. 
The chosen mode is 
\begin{equation*}
\omega = 3.982\, \mathrm{s}^{-1}.
\end{equation*} 
Setting $\Delta \widehat{\alpha}_L = 1 \cdot 10^{-2}$, the amplitudes of the other variables are calculated so that $\Delta \widehat{\v{w}}$ is an  eigenvector corresponding to this mode:
\begin{equation*}
\left(\Delta \widehat{\v{w}} \right)^T = 
\begin{bmatrix}
1.00 \cdot 10^{-2} & 3.99\cdot 10^{-3} & 4.51 \cdot 10^{-3}  & -2.30
\end{bmatrix}.
\end{equation*}
This ensures that the other mode is not present in the initial perturbation, so that we can study the isolated behavior of one mode.
A projection step is then performed in order to make the initial condition satisfy the constraints (see \autoref{sec:discretization}).

\begin{figure}[htbp] 
\centering
\includegraphics[width=0.45\linewidth]{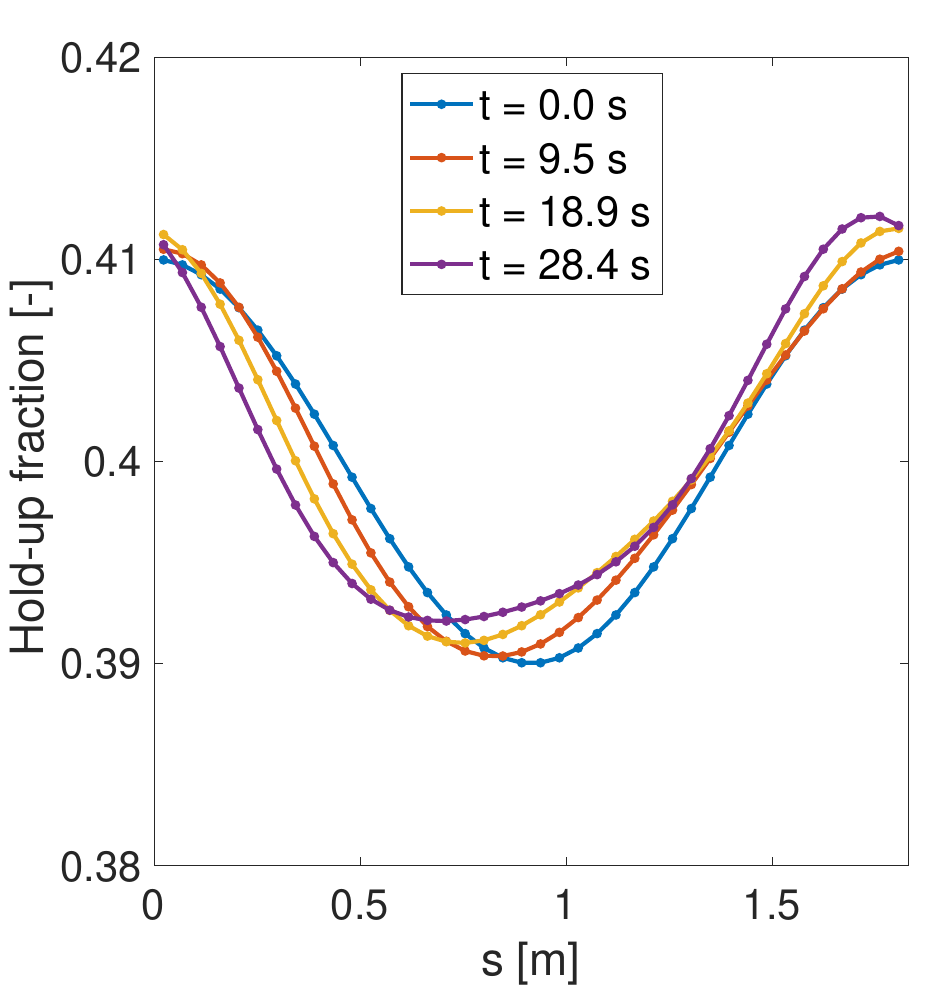} 
\includegraphics[width=0.45\linewidth]{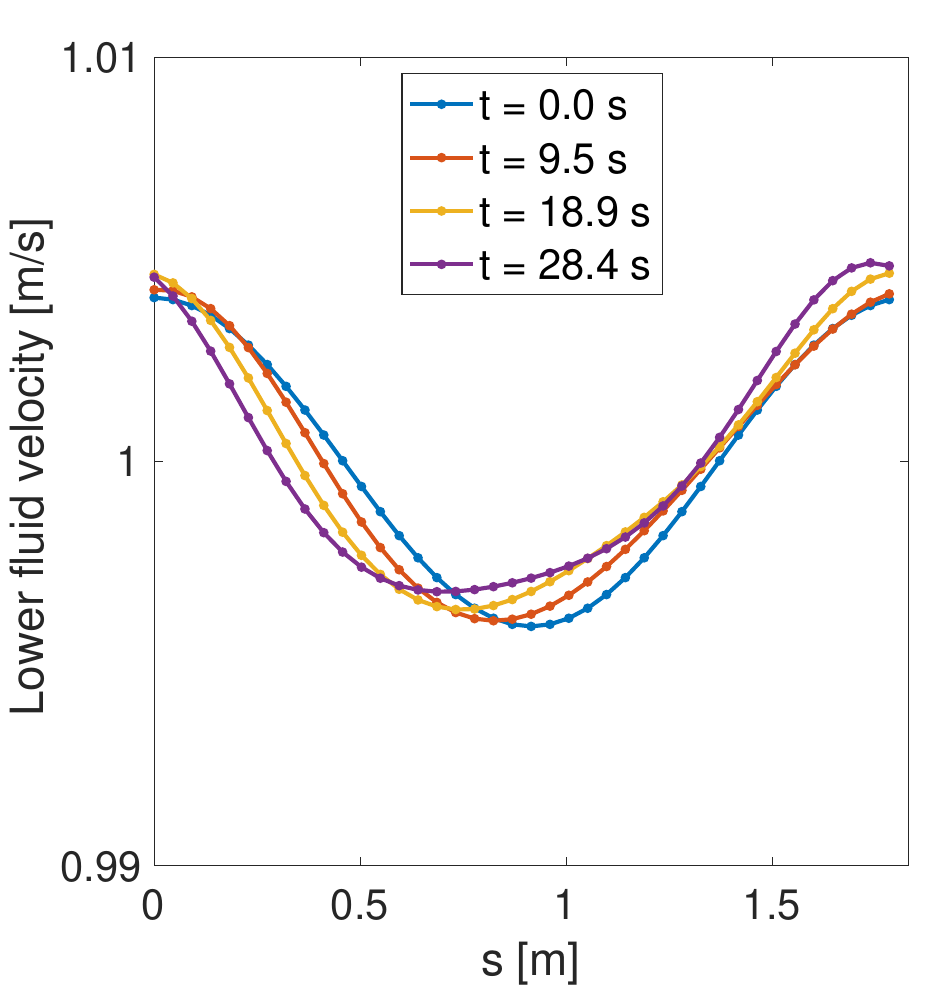} 
\caption{The initial perturbation travels to the right with time. Consecutive snapshots are separated by a time interval of 6 wave periods. Left: lower fluid hold-up. Right: lower fluid velocity.}
\label{fig:results/KH/evolution}
\end{figure}

The initial condition is shown in \autoref{fig:results/KH/evolution}, along with its evolution in time, which is computed up to $t=30 \, \mathrm{s}$.
Setting the initial condition this way yields a wave traveling to the right at velocity $\omega/k=1.16 \, \mathrm{m} \, \mathrm{s}^{-1}$, which remains of approximately constant amplitude since the flow is inviscid and in the well-posed regime, so that $\omega$ has no imaginary component. 
The traveling wave can deform due to the nonlinear character of the governing equations, which is neglected in the linear stability analysis.
This is made apparent by the snapshots of the solution shown in \autoref{fig:results/KH/evolution}, which are separated by an integer number of wave periods: at the time of the last snapshot the wave has traveled through the domain 18 times. 
The solutions do not completely overlap and we see wave steepening taking place.

\begin{figure}[hbtp] 
\centering
\includegraphics[width=0.45\linewidth]{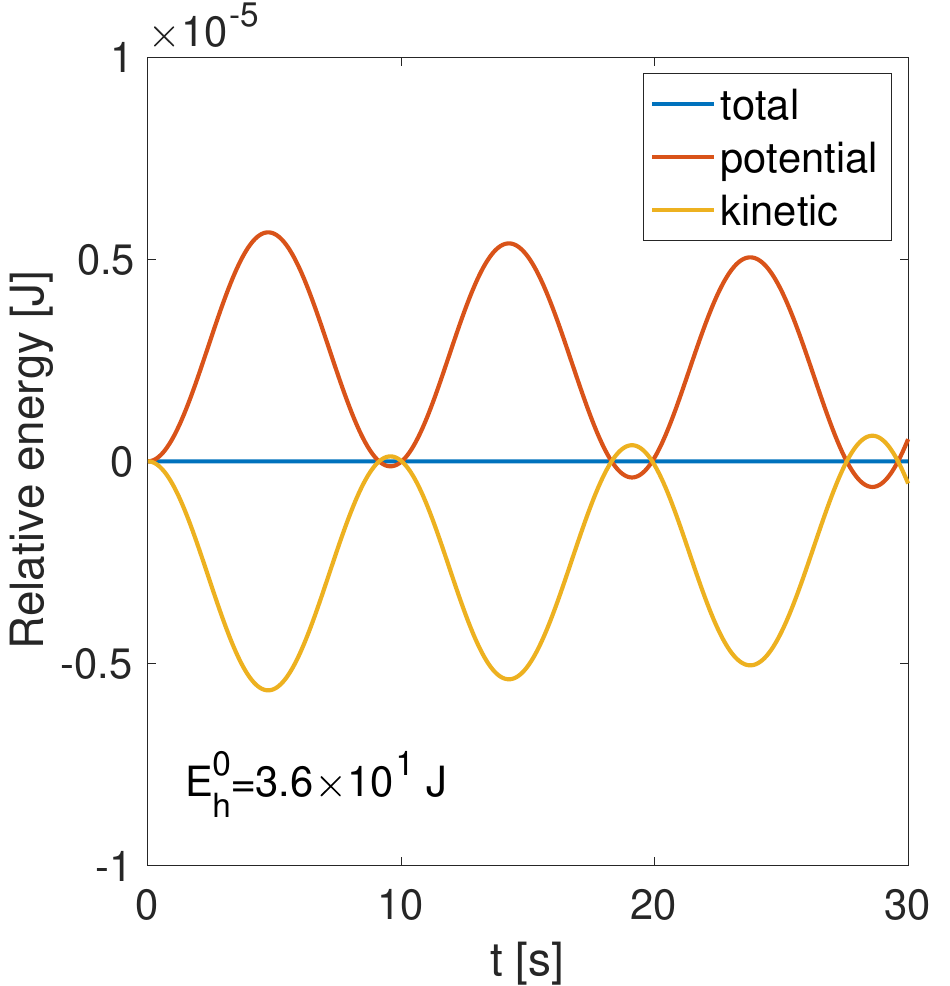} 
\includegraphics[width=0.45\linewidth]{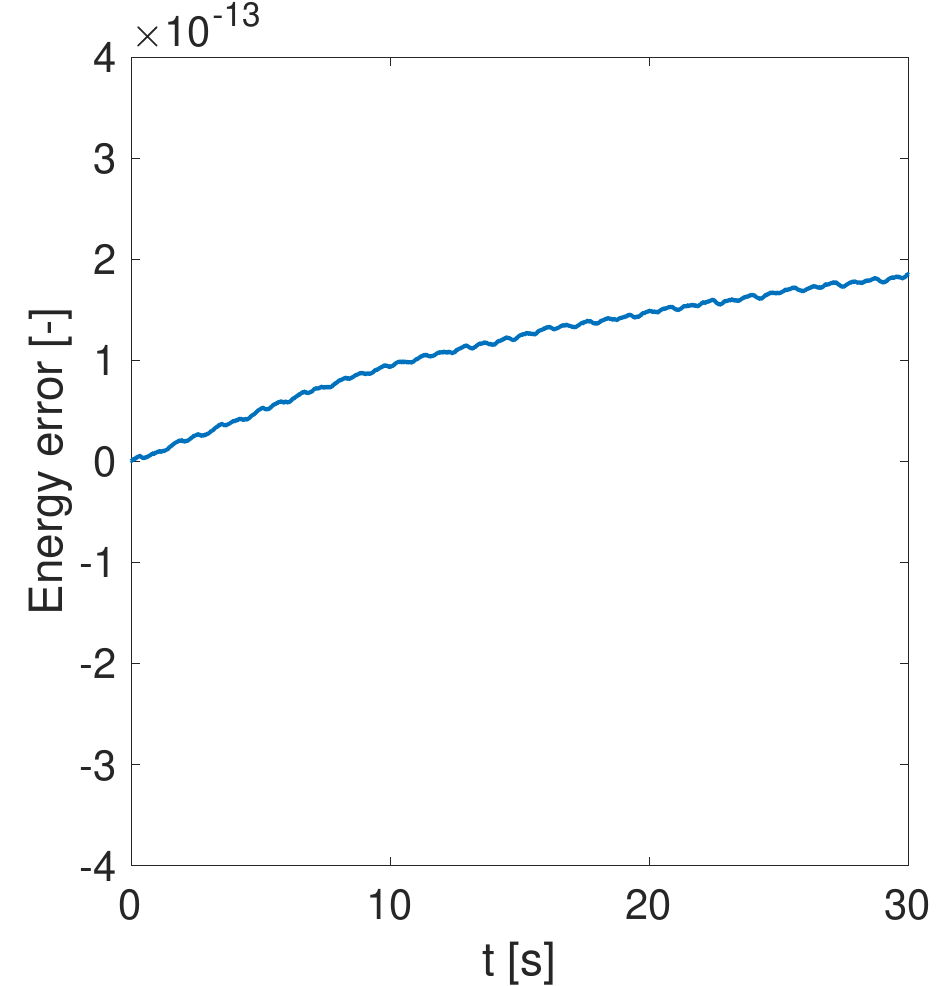}
\caption{Conserved quantities for the traveling wave test case. Left: potential, kinetic and total energy relative to their initial values. Right: $\left(E_{h}-E_h^0\right)/E_h^0$. }
\label{fig:results/KH/conservation_in_time_plot_N_0040}
\end{figure}

\autoref{fig:results/KH/conservation_in_time_plot_N_0040} shows the evolution of the energy. 
In this case, the exchange between kinetic and potential energy is small relative to the total energy of the base state.
This is due to the fact that the wave is roughly constant in time, up to a displacement which does not change the energy.

\begin{figure}[htbp] 
\centering
\includegraphics[width=0.45\linewidth]{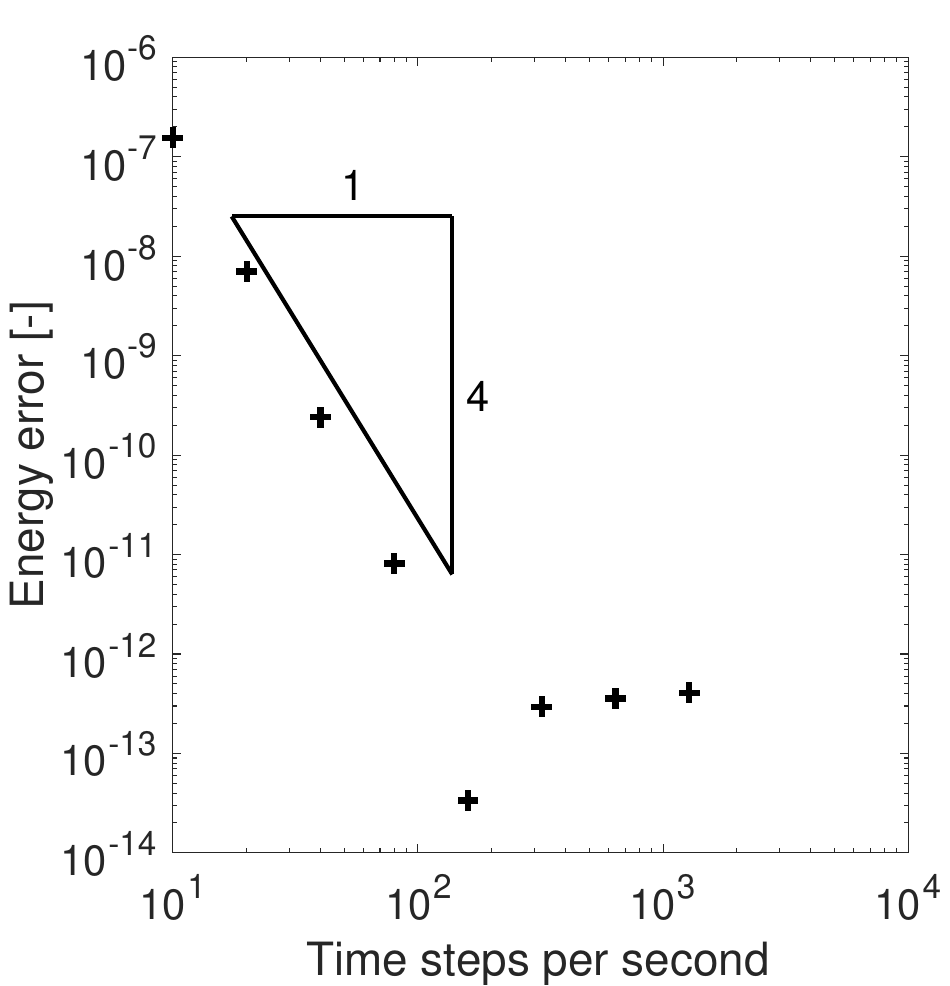} 
\caption{Convergence of the energy error $\left(E_h^{N_t}-E_h^0\right)/E_h^0$ with time step, for the traveling wave test case. 
}
\label{fig:results/KH/E^N-E^0}
\end{figure}

The total energy can again be seen to remain constant up to a high precision.
Like before, this is achieved by using a small time step ($\Delta t = 0.005 \, \mathrm{s}$), with a modest spatial resolution ($N_p = N_u = 40$).
\autoref{fig:results/KH/E^N-E^0} shows how the energy converges with time step refinement. 
The convergence rate is fourth order over a wide range of time steps (matching the order of the time integration method), demonstrating that also for this test case, the spatial discretization conserves energy. 
While the solution moves away from the stable traveling wave predicted by linear analysis, its energy remains constant with time.


\FloatBarrier

 \section{Conclusions} \label{sec:conclusion}

In this article, we have derived the  result that the total mechanical energy (sum of kinetic and potential energy) is a secondary conserved quantity of the incompressible and isothermal TFM. 
This result is in line with the well-known fact that multi-dimensional incompressible frictionless flow equations conserve mechanical energy. Our novel insight is that this conservation statement still holds after averaging: the averaging procedure used to obtain the 1D TFM does not interfere with the energy conservation property.
The approach was based on the formulation of entropy variables and an entropy potential, similar to what is commonly done for the SWE, but with two main differences: (i) we have included a non-conservative pressure term in our analysis, which is shown to be energy-conserving, and (ii) we have obtained our results independent of the duct geometry, which may be a 2D channel or a circular pipe, or any other closed cross-sectional duct shape. 

The second novel result of this paper is a set of numerical fluxes that conserve a discrete form of the mechanical energy. A discretization on a staggered grid was proposed in order to keep the energy conservation property of the non-conservative pressure terms in a discrete sense. 
Although the use of a staggered grid implies that the choice of a discrete energy and entropy potential is not unique, we were able to propose a combination which is such that the discrete analysis is consistent with and analogous to the continuous analysis.
However, one important difference between the continuous and discrete cases remains, namely in the analysis of the level gradient terms (for arbitrary geometries). A geometric relation between the potential energy and the interface height is satisfied exactly in the continuous case, but only approximately in the discrete case. Fortunately, for the specific case of the 2D channel geometry, the condition is satisfied \textit{exactly}, and the discrete level gradient reduces to a form which parallels the continuous form perfectly. For other geometries, such as the pipe, a small numerical energy error persists in the discrete analysis.

Our theoretical derivations are supported by numerical experiments, which show that the proposed energy is indeed exactly conserved by our new spatial discretization in both periodic and closed domains. 
Building on previous work \cite{SanderseVeldman2019}, the discretization also conserves mass and momentum, has strong coupling between momentum and pressure, and is constraint-consistent.
In these experiments the temporal error was negligible (due to a combination of high-order time integration and small time steps), but for future work it is suggested to also make the time integration method energy-conserving \cite{Sanderse2013b}.
Furthermore, the effects of wall friction and pipe inclination need to be added into our formulation. 

Our energy-conserving formulation of the TFM provides a foundation for investigating the nonlinear stability of the model.
For related models, the energy acts as a norm or a convex entropy function of the solution, providing stability bounds, and it should be investigated if the TFM energy has similar implications. 
While we have only considered smooth solutions, in the theory of entropy stability \cite{Tadmor2003}, energy is dissipated at discontinuities.
Therefore, in order to deal with shocks it seems necessary to add suitable diffusion to our formulation, so that the energy becomes strictly decreasing \cite{CastroFjordholmMishraEtAl2013}.

\section*{CRediT}
\textbf{Jurriaan Buist}: Conceptualization, Methodology, Software, Writing - Original Draft; \textbf{Benjamin Sanderse}: Conceptualization, Methodology, Software, Writing - Review \& Editing, Supervision; \textbf{Svetlana Dubinkina}: Writing - Review \& Editing, Supervision; \textbf{Ruud Henkes}: Writing - Review \& Editing, Supervision; \textbf{Kees Oosterlee}: Writing - Review \& Editing, Supervision.

\section*{Funding}

This work was supported by the research program  Shell-NWO/FOM Computational Sciences for Energy Research (CSER), project number 15CSER17, which is partly financed by the Netherlands Organization for Scientific Research (NWO).

\FloatBarrier


\numberwithin{equation}{section}
\numberwithin{figure}{section}
\begin{appendices}

\section{Geometric relations} \label{sec:geometric_details} 
We treat the model equations in a way that is general to arbitrary duct geometries, using general geometric quantities which can be substituted for expressions that are specific to certain duct cross-sectional shapes.
The most important general geometric terms are the $H$-variables, of which we have three for each fluid: $H_U$, $\widehat{H}_U$, $\widetilde{H}_U$, $H_L$, $\widehat{H}_L$, $\widetilde{H}_L$.
We use $H$ (implying something like a height) for each of these variables because they are all invertible functions only of $A_U$ and $A_L$ respectively, and these functions all depend only on the cross-sectional duct shape. 
They are all distinct though, and the relations between these geometric quantities (which hold for arbitrary geometries) are crucial to the results of this paper. 

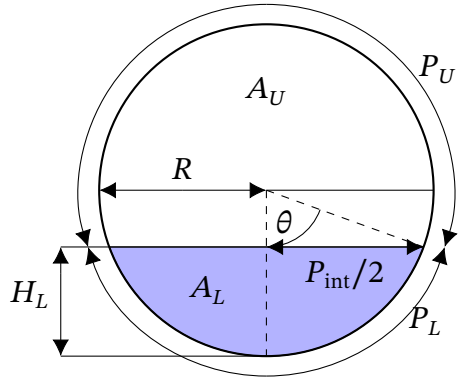
\begin{figure}[!htb]
\begin{center}
\large \centering%
\begin{tikzpicture}[scale=2.2]

\filldraw[fill=blue!30!white, draw=black] (-0.94,-0.342) -- (0.94,-0.342) arc (-20:-160:1) -- cycle;

\draw[thick] (0,0) circle (1);

\node at (-0.35,-0.6) {$A_L$};
\node at (0,0.6) {$A_U$};

\centerarc[-{Latex[width = 2.2mm, length = 2.2mm]}](0,0)(-90:-162.2:1.12);
\centerarc[-{Latex[width = 2.2mm, length = 2.2mm]}](0,0)(-90:-17.8:1.12);
\node[right] at (0.79,-0.79) {$P_L$};

\centerarc[-{Latex[width = 2.2mm, length = 2.2mm]}](0,0)(90:197.8:1.12);
\centerarc[-{Latex[width = 2.2mm, length = 2.2mm]}](0,0)(90:-17.8:1.12);
\node[right] at (0.85,0.73) {$P_U$};

\draw (0,0) -- (1,0);
\draw (0,-0.342) -- (-1.23,-0.342);
\draw (0,-1) -- (-1.23,-1);

\draw[ -{Latex[width = 2.2mm, length = 2.2mm]}] (-1.23, -0.5)--(-1.23,-1);
\draw[ -{Latex[width = 2.2mm, length = 2.2mm]}] (-1.23, -0.5)--(-1.23,-0.342);
\node [left] at (-1.23,-0.63) {$H_L$};


\draw[ -{Latex[width = 2.2mm, length = 2.2mm]}] (-0.5, 0)--(-1,0);
\draw[ -{Latex[width = 2.2mm, length = 2.2mm]}] (-0.5, 0)--(0,0);
\node [above] at (-0.5,0) {$R$};

\draw[dashed] (0,0)--(0.94,-0.342);
\draw[dashed] (0,0)--(0,-1);
\centerarc[](0,0)(-90:-20:0.342);
\node [left] at (0.23,-0.19) {$\theta$};

\draw[ -{Latex[width = 2.2mm, length = 2.2mm]}] (0.5, -0.342)--(0.94,-0.342);
\draw[ -{Latex[width = 2.2mm, length = 2.2mm]}] (0.5, -0.342)--(0,-0.342);
\node [below] at (0.47,-0.342) {$P_\mathrm{int}/2$};

\end{tikzpicture}%
\normalsize \end{center}%
\caption{A schematic of a circular pipe cross-section.}
\label{fig:Saint-Venant/detailed_circular_cross-section_schematic}
\end{figure}

Two geometries of particular interest are the 2D channel and the circular pipe.
For a 2D channel geometry, the following substitutions can be made in the equations: 
\begin{alignat*}{3}
H_L &= A_L,								\quad & \quad  	H_U &= A_U, \nonumber \\
P_L &= 1, 									\quad & \quad 		P_U &= 1, \\
A &= H,										\quad & \quad		P_\mathrm{int} &= 1.	  \nonumber 
\end{alignat*}
For a pipe geometry we have, as in \cite{Akselsen2016},
\begin{alignat*}{3}
H_L &= R(1 - \cos{(\theta)}),								\quad & \quad  		H_U &= R(1 + \cos{(\theta)}), \\
P_L &= 2R \theta, 									\quad & \quad 		P_U &= 2R (\pi - \theta), \\
A &= \pi R^2,										\quad & \quad		P_\mathrm{int} &= 2R \sin{(\theta)}, \\
A_L &= R^2 \left( \theta - \frac{1}{2} \sin{(2 \theta)} \right), 	\quad & \quad A_U &= R^2 \left( \pi - \theta + \frac{1}{2} \sin{(2 \theta)} \right).
\end{alignat*}
In \autoref{fig:Saint-Venant/detailed_circular_cross-section_schematic} we show how the wetted angle $\theta$ is defined.
If $\alpha_L = A_L/A$, then $\pi \alpha_L = \theta - \frac{1}{2} \sin{(2 \theta)}$, and this equation must be solved iteratively in order to obtain $\theta$ from $A_L$, so that the remaining geometric quantities can be calculated.

The integrals \eqref{eq:governing_equations/level_gradient_integrals}
which appear in the governing equations of the two-fluid model are geometry-dependent:
\begin{align}
\widehat{H}_L &\coloneqq  \int_{a_L}   (h-H_L) \, \mathrm{d} a = \int_{0}^{H_L} (h-H_L) w(h) \, \mathrm{d} h, \label{eq:appendix/general_level_gradient_integral/lower} \\
\widehat{H}_U &\coloneqq \int_{a_U}   (h-H_L) \, \mathrm{d} a  = \int_{H_L}^H (h-H_L) w(h) \, \mathrm{d} h, \label{eq:appendix/general_level_gradient_integral/upper}
\end{align}
with $w(h)$ the local width. Note that $w(H_L)=P_\mathrm{int}$.
For a 2D channel geometry, with $A_L = H_L$ and $A_U = H_U$, the width is given by $w(h)=1$ and the integrals evaluate to 
\begin{align*}
\widehat{H}_L &=  -\frac{1}{2} A_L^2, & 
\widehat{H}_U &=  \frac{1}{2}  A_U^2,
\end{align*}
where we have substituted $A_L=A-A_U$. For the pipe geometry, we make the transformation $h=R\left(1 - \cos{(\theta^*)} \right)$, with $\theta^*$ the integration variable and $\theta$ the wetted angle, to get \cite{SanderseSmithHendrix2017}
\begin{align*}
\widehat{H}_L &=  \left[ (R-H_L)A_L - \frac{1}{12} P_\mathrm{int}^3 \right], & 
\widehat{H}_U &= - \left[(R-H_U) A_U - \frac{1}{12} P_\mathrm{int}^3 \right].
\end{align*}
The following derivatives of $\widehat{H}_L$ and $\widehat{H}_U$ are needed in order to calculate $\partial \v{f}/\partial \v{q}$:
\begin{equation}
\d{\widehat{H}_L}{A_L}  = \d{\widehat{H}_L}{H_L} \d{H_L}{A_L},  \quad \quad
\d{\widehat{H}_U}{A_U} = \d{\widehat{H}_U}{H_U} \d{H_U}{A_U}.
\label{eq:appendix/general_derivatives_of_level_gradient_intregrals_to_areas}
\end{equation}
We use Leibniz' rule to calculate
\begin{equation}
\begin{split}
 \d{\widehat{H}_L}{H_L} &= \d{}{H_L} \int_{0}^{H_L} (h-H_L) w(h) \, \mathrm{d} h \\
 &= \left( h(H_L) -H_L \right) w(H_L)\d{H_L}{H_L} - \left( h(0) - H_L \right) w(0) \d{0}{H_L} +  \int_{0}^{H_L}  \d{}{H_L} \left( \left( h - H_L \right) w(h) \right) \, \mathrm{d} h \\
&= - \int_{0}^{H_L} w(h) \, \mathrm{d} h  = -A_L, 
 \end{split}
\end{equation}
and similarly
\begin{equation}
\begin{split}
 \d{\widehat{H}_U}{H_U} &=  \d{}{H_U}  \int_{H_L}^{H}  h w(h) \, \mathrm{d} h =  A_U.
\end{split}
\end{equation}
Substitution in \eqref{eq:appendix/general_derivatives_of_level_gradient_intregrals_to_areas} gives the following relations:
\begin{equation}
\d{\widehat{H}_L}{A_L} = - A_L \d{H_L}{A_L}, \quad  \quad \d{\widehat{H}_U}{A_U} = A_U \d{H_U}{A_U},
\label{eq:appendix/derivative_of_H_hat_relation_to_derivative_of_H}
\end{equation} 
and the inverse of the derivatives appearing on the right-hand sides can also be evaluated using Leibniz' rule:
\begin{equation}
 \d{A_L}{H_L}  = P_\mathrm{int}, \quad \quad
\d{A_U}{H_U} =  P_\mathrm{int}.
\end{equation}

Besides $\widehat{H}_L$ and $\widehat{H}_{U}$, the following geometric quantities are used in \eqref{eq:continuous_energy_conservation/conservative/e_definition_1} and defined as:
 \begin{align}
\widetilde{H}_L &\coloneqq  \int_{a_L}   h \, \mathrm{d} a =  \int_{0}^{H_L} h w(h) \, \mathrm{d} h = \widehat{H}_L + H_L A_L,\label{eq:appendix/general_potential_energy_integral/lower}  \\
\widetilde{H}_U &\coloneqq \int_{a_U}  h \, \mathrm{d} a  = \int_{H_L}^H h w(h) \, \mathrm{d} h = \widehat{H}_U + (H-H_U) A_U, \label{eq:appendix/general_potential_energy_integral/upper}
\end{align}
which can be evaluated by substituting the expressions for $\widehat{H}_L$ and $\widehat{H}_U$.
In order to calculate $\v{v}$ as given by  \eqref{eq:continuous_energy_conservation/v_definition}, we need the derivatives $\mathrm{d}\widetilde{H}_L/A_L$ and $\mathrm{d}\widetilde{H}_U/A_U$. 
They are found by differentiating \eqref{eq:appendix/general_potential_energy_integral/lower} and \eqref{eq:appendix/general_potential_energy_integral/upper}, yielding
\begin{equation}
\d{\widetilde{H}_L}{A_L} = H_L,  \quad \quad
\d{\widetilde{H}_U}{A_U} = H-H_U.
\label{eq:appendix/derivatives_of_potential_energy_intregrals_to_areas}
\end{equation}

\section{Global energy analysis}
\label{sec:global_discrete_energy_conservation}

The main text has described a way to derive the \textit{local} semi-discrete energy conservation equation given by \eqref{eq:semi-discrete_local_energy_conservation/local_energy_conservation_equation}.
In the case of periodic or closed boundaries, this can be integrated in space to yield global energy conservation.
In this section, we directly derive the global energy conservation equation without the intermediate step of the local energy.
This allows us to skip the step of choosing an entropy potential, which means that the derivation will contain less assumptions.
On the other hand, the obtained conditions on the numerical fluxes are not constructive, because they are conditions for the `jumps' of the numerical fluxes, rather than for a single numerical flux at one discrete point. 

The scheme \eqref{eq:discretization/staggered/finite_volume_scheme} described in \autoref{sec:discretization} for a certain pressure volume $i$ and velocity volume $i-1/2$ can be extended to describe the evolution of the entire state vector $\v{q}_{h}$:
\begin{equation}
\d{\v{q}_h}{t} + \v{f}_h + \v{d}_h   = 0,
\label{eq:global_discrete_energy_conservation/global_discrete_model}
\end{equation}
where $\v{q}_{h} =  [q_{1,1} \ldots  q_{1,N},  q_{2,1} \ldots  q_{2,N}, q_{3,1/2}  \ldots q_{3,N-1/2},  q_{4,1/2}  \ldots q_{4,N-1/2}]^T$, and similar expressions for $\v{f}_h$ and $\v{d}_h$. 
For simplicity we only discuss periodic boundary conditions, for which $N_p=N_u=N$.

Similar to the local entropy variable $\v{v}$ we define the global entropy variable
\begin{equation}
\v{v}_h \coloneqq \left[\d{E_{h}}{\v{q}_h}\right]^T.
 \label{eq:global_discrete_energy_conservation/v_vector_definition}
\end{equation}
Taking the inner product of $\v{v}_h$ and \eqref{eq:global_discrete_energy_conservation/global_discrete_model},
the first term yields
\begin{equation*}
\dotp{ \v{v}_h }{ \d{\v{q}_h}{t} }= \d{E_h}{t}.
\end{equation*}
Thus, to obtain global discrete energy conservation, given by $\d{E_h}{t}=0$,
we need the following conditions on $\v{f}_h$ and $\v{d}_h$:
\begin{gather}
\dotp{ \v{v}_h }{\v{f}_h  }= 0,
 \label{eq:global_discrete_energy_conservation/flux_condition} \\
\dotp{ \v{v}_h }{ \v{d}_h } = 0.
 \label{eq:global_discrete_energy_conservation/pressure_condition}
\end{gather}

In order to evaluate $\v{v}_h$, we note that 
\begin{equation}
\pd{E_{h}}{\v{q}_i} = \pd{e_{i-1/2}}{\v{q}_i} + \pd{e_{i+1/2}}{\v{q}_i} =   \v{v}_{i-1/2,i}  + \v{v}_{i+1/2,i} = %
 \begin{bmatrix}
- \frac{1}{2 } \overline{ \left( \frac{q_{3,i}^2}{  \overline{q}_{1,i}^2 } \right) }  +  g_n \left(H-H_{U,i}\right) \\
- \frac{1}{2 } \overline{ \left( \frac{q_{4,i}^2}{  \overline{q}_{2,i}^2 } \right) }  +  g_n H_{L,i}  \\
\frac{q_{3,i-1/2}}{\overline{q}_{1,i-1/2}}  \\
\frac{q_{4,i-1/2}}{\overline{q}_{2,i-1/2}} 
 \end{bmatrix},
 \label{eq:appendix_vh}
\end{equation}
and $\v{v}_h$ follows by assembling this expression for all grid points (ordered by equation, like $\v{q}_h$).
The pressure condition \eqref{eq:global_discrete_energy_conservation/pressure_condition} then evaluates to
\begin{equation*}
\begin{split}
\dotp{\v{v}_h }{ \v{d}_h  }
=  \sum_{i=1}^{N}   Q_{i-1/2}  \left( p_i - p_{i-1} \right) 
=   \sum_{i=1}^{N} \left(  Q_{i+1/2} - Q_{i-1/2}   \right) p_i 
= 0,
\end{split}
\end{equation*}
and is thus satisfied because $Q$ is uniform in space. 

The flux condition  \eqref{eq:global_discrete_energy_conservation/flux_condition} evaluates to
\begin{equation*}
\begin{split}
 \left< \v{v}_h, \v{f}_h  \right>  
  =
 \sum_{i=1}^{N}
  & \left(- \frac{1}{2 } \overline{ \left( \frac{q_{3,i}^2}{  \overline{q}_{1,i}^2 } \right) }  +  g_n \left(H-H_{U,i}\right) \right) \left( f_{1,i+1/2} -  f_{1,i-1/2} \right) 
  + \left( - \frac{1}{2 } \overline{ \left( \frac{q_{4,i}^2}{  \overline{q}_{2,i}^2 } \right) }  +  g_n H_{L,i}   \right) \left( f_{2,i+1/2} - f_{2,i-1/2} \right) \\
  &+ \left( \frac{q_{3,i-1/2}}{\overline{q}_{1,i-1/2}}   \right) \left(  f_{3,i} - f_{3,i-1}  \right) + \left( \frac{q_{4,i-1/2}}{\overline{q}_{2,i-1/2}} \right) \left(  f_{4,i} - f_{4,i-1}  \right). 
  \end{split}
  \end{equation*}
  We split this condition into two conditions: one proportional to $g_n$ and one not proportional to $g_n$:
  \begin{equation*}
   \dotp{ \v{v}_h }{ \v{f}_h }  =  \dotp{ \v{v}_h }{ \v{f}_h }_a  +  \dotp{ \v{v}_h }{ \v{f}_h }_g.
  \end{equation*}
  The advective condition is given by
  \begin{equation*}
\begin{split}
 \dotp{ \v{v}_h }{ \v{f}_h }_a  
= \sum_{i=1}^N \Bigg[
&- \frac{1}{2 } \overline{ \left( \frac{q_{3,i}^2}{  \overline{q}_{1,i}^2 } \right) }  \left( f_{1,i+1/2} -  f_{1,i-1/2} \right) 
  - \frac{1}{2 } \overline{ \left( \frac{q_{4,i}^2}{  \overline{q}_{2,i}^2 } \right) }  \left( f_{2,i+1/2} - f_{2,i-1/2} \right) \\
 & + \left( \frac{q_{3,i-1/2}}{\overline{q}_{1,i-1/2}} \right)  \left(  f_{3,i,a} - f_{3,i-1,a}  \right) + \left( \frac{q_{4,i-1/2}}{\overline{q}_{2,i-1/2}} \right) \left(  f_{4,i,a} - f_{4,i-1,a}  \right) \Bigg].
  \end{split}
  \end{equation*}
Substituting \eqref{eq:semi-discrete_local_energy_conservation/specific_energy_1/f_1_and_f_2} yields
an equation that can be rewritten as
  \begin{equation*}
\begin{split}
 \dotp{ \v{v}_h }{ \v{f}_h }_a  = \sum_{i=1}^N \Bigg[ & \frac{1}{2} \left( \frac{q_{3,i-1/2}^2}{\overline{q}_{1,i-1/2}^2} \frac{q_{3,i-1/2}}{\Delta s} - \frac{q_{3,i+1/2}^2}{\overline{q}_{1,i+1/2}^2} \frac{q_{3,i+1/2}}{\Delta s} \right) +  \frac{1}{2} \left(\frac{q_{4,i-1/2}^2}{\overline{q}_{2,i-1/2}^2} \frac{q_{4,i-1/2}}{\Delta s} -  \frac{q_{4,i+1/2}^2}{\overline{q}_{2,i+1/2}^2} \frac{q_{4,i+1/2}}{\Delta s} \right) \\ 
 + &\left(\frac{q_{3,i+1/2}}{  \overline{q}_{1,i+1/2} } - \frac{q_{3,i-1/2}}{  \overline{q}_{1,i-1/2} } \right) \overline{\left(  \frac{q_{3,i} }{  \overline{q}_{1,i}}  \right)}  \frac{\overline{q}_{3,i-1/2}}{\Delta s} 
    + \left( \frac{q_{3,i-1/2}}{\overline{q}_{1,i-1/2}}   \right) \left(  f_{3,i,a} - f_{3,i-1,a}  \right)    \\
   +&  \left(\frac{q_{4,i+1/2}}{  \overline{q}_{2,i+1/2} } - \frac{q_{4,i-1/2}}{  \overline{q}_{2,i-1/2} } \right) \overline{ \left(  \frac{q_{4,i} }{ \overline{q}_{2,i}} \right) }  \frac{\overline{q}_{4,i}}{\Delta s}  
   + \left( \frac{q_{4,i-1/2}}{\overline{q}_{2,i-1/2}} \right) \left(  f_{4,i,a} - f_{4,i-1,a}  \right) \Bigg].
      \end{split}
  \end{equation*}
  Here, the sum over the entries on the first lines evaluates to zero, since each term has a matching term of opposite sign and index shifted by 1 (even the boundary terms, in case of periodic boundaries).
  In order for this to also hold for the terms in the second and third lines, we need to satisfy the condition
    \begin{align*}
  \left(\frac{q_{3,i+1/2}}{  \overline{q}_{1,i+1/2} } - \frac{q_{3,i-1/2}}{  \overline{q}_{1,i-1/2} } \right) \overline{\left(  \frac{q_{3,i} }{  \overline{q}_{1,i}}  \right)}  \frac{\overline{q}_{3,i-1/2}}{\Delta s} +  \left( \frac{q_{3,i-1/2}}{\overline{q}_{1,i-1/2}}   \right)  f_{3,i,a}  =  \left( \frac{q_{3,i+1/2}}{\overline{q}_{1,i+1/2}}   \right)  f_{3,i,a},
   \end{align*}
   and similar for $f_{4,i,a}$. These are indeed satisfied with our choice \eqref{eq:semi-discrete_local_energy_conservation/specific_energy_1/f_3_i_a}. 


The condition proportional to $g_n$, after substitution of \eqref{eq:semi-discrete_local_energy_conservation/specific_energy_1/f_1_and_f_2},  is given by 
      \begin{equation*}
\begin{split}
 \dotp{ \v{v}_h }{ \v{f}_h }_g  
    = \sum_{i=1}^N \Bigg[ & g_n \left(H-H_{U,i}\right)  \left( \frac{ q_{3,i+1/2} }{\Delta s}-  \frac{ q_{3,i-1/2} }{\Delta s}\right) 
  +   g_n H_{L,i}   \left( \frac{ q_{4,i+1/2} }{\Delta s}- \frac{ q_{4,i-1/2} }{\Delta s} \right) \\
   + & g_n \left( \frac{q_{3,i-1/2}}{\overline{q}_{1,i-1/2}}   \right) \left(  f_{3,i,g} - f_{3,i-1,g}  \right) 
   + g_n \left( \frac{q_{4,i-1/2}}{\overline{q}_{2,i-1/2}} \right) \left(  f_{4,i,g} - f_{4,i-1,g}   \right) \Bigg],
    \end{split}
  \end{equation*}
  and it can be rewritten as
      \begin{equation*}
\begin{split}
 \dotp{ \v{v}_h }{ \v{f}_h }_g = \sum_{i=1}^N \Bigg[
     &g_n\frac{  \Delta s \left(  f_{3,i,g} - f_{3,i-1,g}  \right) -\left( H- H_{U,i} \right)  \left(q_{1,i-1} + q_{1,i}  \right)}{\overline{q}_{1,i-1/2}}  \frac{q_{3,i-1/2}}{\Delta s}  + g_n \left( H- H_{U,i} \right) \frac{q_{3,i+1/2}}{\Delta s}    \\
     + &g_n \frac{  \Delta s \left(   f_{4,i,g} - f_{4,i-1,g}  \right) - H_{L,i}   \left(q_{2,i-1} + q_{2,i}  \right)}{\overline{q}_{2,i-1/2}}  \frac{q_{4,i-1/2}}{\Delta s}  + g_n H_{L,i} \frac{q_{4,i+1/2}}{\Delta s}\Bigg].     
  \end{split}
  \end{equation*}
  \normalsize
  Now, in order for this to be conservative, we need the first term in each line to be equal but opposite in sign to the second term in each line (shifted in index by 1).
This yields the following conditions:
\begin{equation}
\llbracket f_{3,i-1/2,g} \rrbracket = -  \frac{ \overline{q}_{1,i-1/2} }{\Delta s} \llbracket H_{U,i-1/2} \rrbracket, 
\qquad
\llbracket  f_{4,i-1/2,g} \rrbracket =  \frac{  \overline{q}_{2,i-1/2} }{\Delta s} \llbracket  H_{L,i-1/2} \rrbracket,
\label{eq:global_discrete_energy_conservation/geometric_condition_lower_a}
\end{equation}
which upon substitution of  \eqref{eq:semi-discrete_local_energy_conservation/specific_energy_1/channel_result_f_4_i_g} reduce to the geometric conditions \eqref{eq:semi-discrete_local_energy_conservation/specific_energy_1/check_of_better_deduction/geometric_condition/together}.

In conclusion, the results of the global discrete analysis are consistent with our local discrete analysis. The additional insight from the global analysis is that the geometric conditions \eqref{eq:global_discrete_energy_conservation/geometric_condition_lower_a} or 
 \eqref{eq:semi-discrete_local_energy_conservation/specific_energy_1/check_of_better_deduction/geometric_condition/together} are \textit{independent of the choice of the entropy potential}. 
 This confirms that the choice of entropy potential does not limit the results.

\rmk{The global energy analysis can also be performed without requiring interpolation of the potential energy to the velocity grid points, as needed in the definition of $e_{i-1/2}$ given by \eqref{eq:semi-discrete_local_energy_conservation/specific_energy_1/energy_definition}. Instead, one can directly define
\begin{equation}
    E_{h} = \sum_{i=1}^{N_p} \left( \rho_U g_n \widetilde{H}_{U,i} \Delta s +  \rho_L g_n \widetilde{H}_{L,i} \Delta s  \right)  + \sum_{i=1}^{N_u}
    \left( \frac{1}{2} \frac{q_{3,i-1/2}^2}{ \overline{q}_{1,i-1/2}  } + \frac{1}{2} \frac{q_{4,i-1/2}^2}{ \overline{q}_{2,i-1/2}} \right).
\end{equation}
It can be verified that this leads to the same $\v{v}_{h}$ as given by \eqref{eq:appendix_vh}, and consequently the geometric condition \eqref{eq:global_discrete_energy_conservation/geometric_condition_lower_a} remains present.}

\end{appendices}

\bibliography{library_bibtex}      

\begin{thebibliography}{10}

\bibitem{AbgrallKarni2009}
R.~Abgrall and S.~Karni.
\newblock Two-layer shallow water system: {{A}} relaxation approach.
\newblock {\em SIAM Journal on Scientific Computing}, 31(3):1603--1627, 2009.

\bibitem{Akselsen2016}
A.~H. Akselsen.
\newblock {\em Efficient {{Numerical Methods}} for {{Waves}} in
  {{One}}-{{Dimensional Two}}-{{Phase Pipe Flows}}}.
\newblock PhD thesis, Norwegian University of Science and Technology, 2016.

\bibitem{AursandHammerMunkejordEtAl2013}
P.~Aursand, M.~Hammer, S.~T. Munkejord, and {\O}.~Wilhelmsen.
\newblock Pipeline transport of {{CO2}} mixtures: {{Models}} for transient
  simulation.
\newblock {\em International Journal of Greenhouse Gas Control}, 15:174--185,
  2013.

\bibitem{BerryZouZhaoEtAl2014}
R.~A. Berry, L.~Zou, H.~Zhao, H.~Zhang, J.~W. Peterson, R.~C. Martineau, S.~Y.
  Kadioglu, and D.~Andrs.
\newblock {{RELAP}}-7 {{Theory Manual}}.
\newblock Technical Report INL/EXT-14-31366, {Idaho National Laboratory}, 2014.

\bibitem{BuistSandersevanHalderEtAl2019}
J.~Buist, B.~Sanderse, Y.~{van Halder}, B.~Koren, and G.~J. {van Heijst}.
\newblock Machine learning for closure models in multiphase flow applications.
\newblock In {\em Proceedings of the 3rd {{International Conference}} on
  {{Uncertainty Quantification}} in {{Computational Sciences}} and
  {{Engineering}} ({{UNCECOMP}} 2019)}, pages 379--399, {Crete, Greece}, 2019.

\bibitem{CastroFjordholmMishraEtAl2013}
M.~J. Castro, U.~S. Fjordholm, S.~Mishra, and C.~Par{\'e}s.
\newblock Entropy conservative and entropy stable schemes for nonconservative
  hyperbolic systems.
\newblock {\em SIAM Journal on Numerical Analysis}, 51(3):1371--1391, 2013.

\bibitem{Churchill1977}
S.~W. Churchill.
\newblock Friction factor equation spans all fluid flow regimes.
\newblock {\em Chemical Engineering}, 84:91--92, 1977.

\bibitem{CoppolaCapuanodeLuca2019}
G.~Coppola, F.~Capuano, and L.~{de Luca}.
\newblock Discrete energy-conservation properties in the numerical simulation
  of the {{Navier}}\textendash{{Stokes}} equations.
\newblock {\em Applied Mechanics Reviews}, 71(1), 2019.

\bibitem{FjordholmMishraTadmor2009}
U.~Fjordholm, S.~Mishra, and E.~Tadmor.
\newblock Energy {{Preserving}} and {{Energy Stable Schemes}} for the {{Shallow
  Water Equations}}.
\newblock In F.~Cucker, A.~Pinkus, and M.~J. Todd, editors, {\em Foundations of
  {{Computational Mathematics}}, {{Hong Kong}} 2008}, pages 93--139. {Cambridge
  University Press}, 2009.

\bibitem{Fjordholm2012}
U.~S. Fjordholm.
\newblock Energy {{Conservative}} and {{Stable Schemes}} for the {{Two}}-layer
  {{Shallow Water Equations}}.
\newblock In {\em Hyperbolic {{Problems}}}, volume 17 \& 18 of {\em Series in
  {{Contemporary Applied Mathematics}}}, pages 414--421. {Co-Published with
  Higher Education Press}, 2012.

\bibitem{FjordholmMishraTadmor2011}
U.~S. Fjordholm, S.~Mishra, and E.~Tadmor.
\newblock Well-balanced and energy stable schemes for the shallow water
  equations with discontinuous topography.
\newblock {\em Journal of Computational Physics}, 230(14):5587--5609, 2011.

\bibitem{GoldszalDanielsonBansalEtAl2007}
A.~Goldszal, T.~J. Danielson, K.~M. Bansal, Z.~L. Yang, S.~T. Johansen, and
  G.~Depay.
\newblock {{LedaFlow 1D}}: {{Simulation}} results with multiphase
  gas/condensate and oil/gas field data.
\newblock In {\em {{BHR Group}} - 13th {{International Conference}} on
  {{Multiphase Production Technology}}}, page~15, 2007.

\bibitem{Holmas2010}
H.~Holm{\aa}s.
\newblock Numerical simulation of transient roll-waves in two-phase pipe flow.
\newblock {\em Chemical Engineering Science}, 65(5):1811--1825, 2010.

\bibitem{Ishii1975}
M.~Ishii.
\newblock {\em Thermo-Fluid Dynamic Theory of Two-Phase Flow}.
\newblock {Eyrolles}, {Paris}, 1975.

\bibitem{IshiiMishima1984}
M.~Ishii and K.~Mishima.
\newblock Two-fluid model and hydrodynamic constitutive relations.
\newblock {\em Nuclear Engineering and Design}, 82(2-3):107--126, 1984.

\bibitem{Keyfitz2001}
B.~L. Keyfitz.
\newblock Mathematical properties of nonhyperbolic models for incompressible
  two-phase flow.
\newblock In {\em Proceedings of the {{ICMF}}}, 2001.

\bibitem{KreissYstrom2006}
H.-O. Kreiss and J.~Ystr{\"o}m.
\newblock A note on viscous conservation laws with complex characteristics.
\newblock {\em BIT Numerical Mathematics}, 46(S1):55--59, 2006.

\bibitem{Leveque2002}
R.~J. Leveque.
\newblock {\em Finite {{Volume Methods}} for {{Hyperbolic Problems}}}.
\newblock {Cambridge University Press}, 2002.

\bibitem{LiaoMeiKlausner2008}
J.~Liao, R.~Mei, and J.~F. Klausner.
\newblock A study on the numerical stability of the two-fluid model near
  ill-posedness.
\newblock {\em International Journal of Multiphase Flow}, 34(11):1067--1087,
  2008.

\bibitem{LopezdeBertodanoFullmerClausseEtAl2017}
M.~{L{\'o}pez de Bertodano}, W.~Fullmer, A.~Clausse, and V.~H. Ransom.
\newblock {\em Two-{{Fluid Model Stability}}, {{Simulation}} and {{Chaos}}}.
\newblock {Springer International Publishing}, {Cham, Switzerland}, 2017.

\bibitem{LopezdeBertodanoFullmerClausse2016}
M.~{Lopez de Bertodano}, W.~D. Fullmer, and A.~Clausse.
\newblock One-dimensional two-fluid model for wavy flow beyond the
  {{Kelvin}}\textendash{{Helmholtz}} instability: {{Limit}} cycles and chaos.
\newblock {\em Nuclear Engineering and Design}, 310:656--663, 2016.

\bibitem{LyczkowskiGidaspowSolbrigEtAl1978}
R.~W. Lyczkowski, D.~Gidaspow, C.~W. Solbrig, and E.~D. Hughes.
\newblock Characteristics and stability analyses of transient one-dimensional
  two-phase flow equations and their finite difference approximations.
\newblock {\em Nuclear Science and Engineering}, 66(3):378--396, 1978.

\bibitem{Montini2011}
M.~Montini.
\newblock {\em Closure Relations of the One-Dimensional Two-Fluid Model for the
  Simulation of Slug Flows}.
\newblock PhD thesis, Imperial College London, 2011.

\bibitem{Munkejord2006}
S.~T. Munkejord.
\newblock {\em Analysis of the {{Two}}-{{Fluid Model}} and the {{Drift}}-{{Flux
  Model}} for {{Numerical Calculation}} of {{Two}}-{{Phase Flow}}}.
\newblock PhD thesis, Norwegian University of Science and Technology,
  {Trondheim}, 2006.

\bibitem{OckendonOckendon2017}
H.~Ockendon and J.~R. Ockendon.
\newblock How to mitigate sloshing.
\newblock {\em SIAM Review}, 59(4):905--911, 2017.

\bibitem{Sanderse2013b}
B.~Sanderse.
\newblock Energy-conserving {{Runge}}\textendash{{Kutta}} methods for the
  incompressible {{Navier}}\textendash{{Stokes}} equations.
\newblock {\em Journal of Computational Physics}, 233:100--131, 2013.

\bibitem{SanderseSmithHendrix2017}
B.~Sanderse, I.~E. Smith, and M.~H.~W. Hendrix.
\newblock Analysis of time integration methods for the compressible two-fluid
  model for pipe flow simulations.
\newblock {\em International Journal of Multiphase Flow}, 95:155--174, 2017.

\bibitem{SanderseVeldman2019}
B.~Sanderse and A.~E.~P. Veldman.
\newblock Constraint-consistent {{Runge}}\textendash{{Kutta}} methods for
  one-dimensional incompressible multiphase flow.
\newblock {\em Journal of Computational Physics}, 384:170--199, 2019.

\bibitem{StewartWendroff1984}
H.~B. Stewart and B.~Wendroff.
\newblock Two-phase flow: {{Models}} and methods.
\newblock {\em Journal of Computational Physics}, 56(3):363--409, 1984.

\bibitem{Tadmor2003}
E.~Tadmor.
\newblock Entropy stability theory for difference approximations of nonlinear
  conservation laws and related time-dependent problems.
\newblock {\em Acta Numerica}, 12:451--512, 2003.

\bibitem{TadmorZhong2008}
E.~Tadmor and W.~Zhong.
\newblock Energy-{{Preserving}} and {{Stable Approximations}} for the
  {{Two}}-{{Dimensional Shallow Water Equations}}.
\newblock In H.~{Munthe-Kaas} and B.~Owren, editors, {\em Mathematics and
  {{Computation}}, a {{Contemporary View}}}, volume~3, pages 67--94. {Springer
  Berlin Heidelberg}, 2008.

\bibitem{Thorpe1969}
S.~A. Thorpe.
\newblock Experiments on the instability of stratified shear flows: Immiscible
  fluids.
\newblock {\em Journal of Fluid Mechanics}, 39(1):25--48, 1969.

\bibitem{vantHofVeldman2012}
B.~{van't Hof} and A.~E.~P. Veldman.
\newblock Mass, momentum and energy conserving ({{MaMEC}}) discretizations on
  general grids for the compressible {{Euler}} and shallow water equations.
\newblock {\em Journal of Computational Physics}, 231(14):4723--4744, 2012.

\bibitem{Wallis1969}
G.~B. Wallis.
\newblock {\em One-Dimensional Two-Phase Flow}.
\newblock {McGraw-Hill}, {New York}, 1969.

\end{thebibliography}
\bibliographystyle{abbrv}

\end{document}